%% file: Draft13.tex
\documentclass[a4 paper, 11pt]{article}
\pdfoutput=1
\usepackage{comment}
\usepackage{jheppub}
\usepackage{subcaption}
\usepackage{amsmath}
\usepackage{mathrsfs}
\usepackage{amssymb}
\usepackage{amscd}
\usepackage{dsfont}
\usepackage{enumerate}
\usepackage{amsfonts}
\usepackage{epsfig}
\usepackage{mathtools}
\usepackage{yfonts}
\usepackage{bbold}
\usepackage{breqn}
\usepackage[utf8]{inputenc}
\usepackage[english]{babel}
\usepackage{graphicx}
\usepackage{empheq}
\usepackage{relsize}

\usepackage{cancel}

\input{preamble.tex}

\newcommand{\z}{\left}
\newcommand{\y}{\right}
\newcommand{\p}{\partial}
\newcommand{\tMOh}{\tilde{\MO}_{h,\bar{h}}}

\title{
Symmetries and operators in $T\bar{T}$ deformed CFTs 
}

\author{Liangyu Chen$^\gamma$,} 

\author{Zhengyuan Du$^{\gamma,\tau,\sigma}$,}
\author{Kangning Liu$^{\gamma,\tau}$,}
\author{Wei Song$^{\gamma,\tau, \eta}$}

\emailAdd{liangyu-chen@mail.tsinghua.edu.cn} 
\emailAdd{duzy21@mails.tsinghua.edu.cn}
\emailAdd{lkn22@mails.tsinghua.edu.cn}
\emailAdd{wsong2014@mail.tsinghua.edu.cn}

\abstract
{ 
$T\bar{T}$-deformed CFTs are known to possess nonlocal {Virasoro$\times$Virasoro symmetries}{}, whose currents depend on integration of the stress tensor. Unfortunately, these symmetries do not act in a tractable manner on the undeformed operators, which are defined by evolving local operators in the undeformed CFT on a constant-time slice using the deformed Hamiltonian. In this paper, we explicitly construct two distinct classes of operators: (i) dressed operators, which are primary with respect to the {Virasoro$\times$Virasoro symmetries}{}, and (ii) physical operators, a new type of operator that depends on fundamental fields at specific points and is therefore local at the classical level. While dressed operators preserve the conformal symmetry structure, they are themselves nonlocal in the sense that they depend on integrated quantities. Physical operators, by contrast, depend only on local fields. Moreover, physical operators can be expressed in terms of dressed operators, and this relation can be promoted to a definition of such operators at the quantum level. This enables the use of {Virasoro$\times$Virasoro symmetries}{} to systematically compute physical quantities such as correlation functions. We calculate the two-point correlation functions of these operators in momentum space and find that our results align with both string theory predictions and field theory calculations. Additionally, we explore the relationship between physical operators and alternative operator definitions proposed in the literature.
}

\begin{document}
\maketitle

\section{Introduction}
$T\bar{T}$ deformations~\cite{Zamolodchikov:2004ce} exhibit two remarkable properties: they are simultaneously irrelevant and solvable. 
Unlike generic irrelevant deformations, which are typically intractable, $T\bar{T}$-deformed theories remain analytically controllable and lead to novel UV behavior. 
This UV structure manifests in a complex spectrum or Hagedorn growth, signaling the emergence of nonlocality. 
Remarkably, the deformed spectrum on the cylinder admits a compact expression in terms of the deformation parameter and the undeformed spectrum~\cite{Smirnov:2016lqw,Cavaglia:2016oda}.
Furthermore, key properties of the original theory persist under the deformation, including integrability~\cite{Smirnov:2016lqw}, modular invariance~\cite{Datta:2018thy,Aharony:2018bad}, and universal torus partition function behavior~\cite{Apolo:2023aho}. 
Scattering amplitudes~\cite{Dubovsky:2012wk,Dubovsky:2013ira,Dubovsky:2017cnj} and partition functions~\cite{Cardy:2018sdv} can likewise be constructed through dressing of their undeformed counterparts.

The combination of irrelevance and solvability makes $T\bar{T}$ deformation a powerful tool for constructing tractable holographic dualities beyond standard AdS/CFT. While irrelevant deformations of CFTs typically correspond to modifying the asymptotic boundary conditions in the bulk, the solvability of $T\bar{T}$ ensures the resulting duality remains computationally manageable.
In the context of AdS$_3$/CFT$_2$, the double-trace nature of $T\bar{T}$ deformation leaves the bulk solution unchanged but introduces mixed boundary conditions\footnote{Such mixed boundary conditions also arise for more general stress-tensor deformations at the classical level \cite{Ran:2025xas}.} for gravity in asymptotically AdS$_3$ spacetimes~\cite{Guica:2019nzm}. For pure Einstein gravity, this is equivalent to either imposing a finite radial cutoff~\cite{McGough:2016lol} or attaching an additional AdS$_3$ patch to the asymptotic boundary~\cite{Apolo:2023vnm}.

Non-AdS holography with more explicit modification of the bulk spacetime can be constructed in type IIB string theory with NS-NS flux {\cite{Giveon:2017nie,Kutasov:2017myj,Asrat:2017tzd,Apolo:2019zai}}{}. At a generic value of the NS-NS magnetic charge $k$, spectrum of string theory on $AdS_3\times \mathcal N$ contains the short string sector and the long string sector \cite{Maldacena:2000hw}. In the tensionless limit, i.e., superstring at $k=1$ \citep{Eberhardt:2018ouy} or bosonic string at $k=3$ \cite{Eberhardt:2025sbi}, the short string sector disappears, and the holographic dual is described by a symmetric product CFT.
    Away from the tensionless point, the complete holographic dual theory is a marginal deformation of the symmetric product structure {\cite{Eberhardt:2021vsx,Balthazar:2021xeh,Martinec:2021vpk}}{}. Nevertheless, the spectrum of the long string sector can still be described by a symmetric product structure \cite{Eberhardt:2019qcl}. By performing a $T\bar T$ deformation to the seed CFT of a symmetric product CFT, one can construct the single-trace $T\bar T$ deformed CFTs \cite{Giveon:2017nie,Apolo:2023aho,Benjamin:2023nts}, which are related holographically to the long string sector of an asymptotically linear dilaton background \cite{Giveon:2017nie}. 
Holographic interpretations of the single-trace $T\bar T$ deformation include the decoupling limit of black strings \cite{Giveon:2017nie,Chakraborty:2020swe,Chakraborty:2023mzc,Chakraborty:2023zdd}, or the so-called TsT/${T\bar T}$ correspondence \cite{Apolo:2018qpq, Araujo:2018rho, Borsato:2018spz, Apolo:2019zai,Cui:2023jrb} \footnote{
The dual string background can be obtained by performing a TsT transformation, which is a solution-generating technique to generate new string backgrounds \cite{Lunin:2005jy,Maldacena:2008wh}, to AdS$_3\times \mathcal N$ background supported by NS-NS flux \cite{Apolo:2018qpq, Araujo:2018rho, Borsato:2018spz, Apolo:2019zai}. A crucial difference in the two proposals is that the deformation parameter of $T\bar T$ deformation in the decoupling limit proposal is identified with the string tension, while it is identified with the TsT parameter in the TsT/${T\bar T}$ proposal. At zero temperature, the bulk solutions in the two proposals agree exactly with each other. At finite temperature, the choice of the B-field are different in the two proposals. Nevertheless the string spectrum and black hole thermodynamics can be worked out consistently in both proposals. }.   
    So far, single-trace $T\bar T$ deformation has only been explicitly defined for symmetric product CFT$_2$, and hence the holographic duality is expected to be complete only in the tensionless limit \cite{Dei:2024sct}.  At a generic value of $k$, however, we still expect correspondence between the long string sector in the bulk theory and the single-trace $T\bar T$-deformed CFT, at least for some quantities \footnote{It is an interesting open question to reproduce results in the short string sector \cite{Giribet:2017imm,Giveon:2017nie,Fichet:2023xbu} from the dual field theory.}. Evidence for the single-trace $T\bar T$ holography
includes matching of the long string spectrum \cite{Kutasov:2017myj, Apolo:2019zai} and the spectrum of the twisted sector \cite{Apolo:2023aho}, black hole entropy \cite{Apolo:2019zai,Chakraborty:2020swe,Apolo:2021wcn,Chakraborty:2023mzc,Chang:2023kkq,Chakraborty:2023zdd}, and symmetries \cite{Guica:2022gts,Georgescu:2022iyx, Du:2024bqk, Chakraborty:2023wel}. Notably, the worldsheet derivation of two-point functions at finite deformation parameter~\cite{Cui:2023jrb} can be reproduced by the field theory computations~\cite{Aharony:2023dod} in the high energy limit. 
For further developments in $T\bar{T}$ deformations and their holographic realizations, see~\cite{Jiang:2023rxa,He:2025ppz,Brizio:2024arr} and references therein.

Given that $T\bar{T}$ deformations are challengingly irrelevant yet remain solvable, one may wonder whether symmetries play a fundamental role in such theories. Indeed, when a CFT is deformed by $T\bar{T}$, {Virasoro$\times$Virasoro symmetries}{} persists, but its generators become nonlocal. This has been demonstrated from multiple perspectives: in field theory through Lagrangian~\cite{Guica:2020uhm} and Hamiltonian approaches~\cite{Guica:2022gts,Monten:2024efe}, in gravity~\cite{Guica:2019nzm,Kraus:2021cwf,Georgescu:2022iyx}, and in the string worldsheet of the TsT-transformed $\text{AdS}$ background~\cite{Du:2024tlu,Du:2024bqk}.  

Given these symmetries, a natural question arises: How can they be used to organize states or operators, and how might they simplify the computation of physical quantities? A particularly important application lies in calculating correlation functions, which have historically been challenging. Perturbative studies have been explored in~\cite{Kraus:2018xrn,Chen:2018eqk,He:2019vzf,He:2020udl,He:2023kgq}, while non-perturbative progress includes a Callan-Symanzik equation derived in~\cite{Cardy:2019qao}, an exact momentum-space result obtained from the string worldsheet via the TsT/$T\bar{T}$ correspondence~\cite{Cui:2023jrb}, and a large-momentum analysis using path integral methods~\cite{Aharony:2023dod,Barel:2024dgv}. Notably, the two-point function in the untwisted sector of~\cite{Cui:2023jrb} and the large-momentum result of~\cite{Aharony:2023dod} agree precisely, and both satisfy the Callan-Symanzik equation of~\cite{Cardy:2019qao}. However, the details of these results differ, and it remains unclear whether the operators considered in these works are equivalent.  
The primary objective of this work is to leverage {Virasoro$\times$Virasoro symmetries}{} to motivate the definition of operators, enabling systematic and symmetry-guided computations of correlation functions.

In this paper, we study $T\bar{T}$-deformed CFTs within the Hamiltonian formalism~\cite{Kruthoff:2020hsi}. The deformation on the plane induces a canonical flow that mixes phase space variables through a Bogoliubov transformation. The deformed Hamiltonian obeys the flow equation
\begin{align}\label{Hflowintro}
    \mathcal{D}_\lambda H = 0, \quad \mathcal{D}_\lambda \equiv \partial_\lambda - i[\mathcal{X},\cdot]
\end{align}
where $\mathcal{X}$ is a bilocal integral constructed from the stress tensor.

Starting from an undeformed operator $\mathcal{O}^{(0)}(0,y) = \mathcal{O}^{\text{CFT}}(0,y)$ on the initial time slice $t=0$, two types
of operators have been discussed in the literature:
\begin{itemize}
    \item {\it Undeformed} operators: Defined via time evolution of initial operators
    \begin{align}\label{udeq}
        \mathcal{O}^{(0)}(t,y) = e^{iHt}\mathcal{O}^{(0)}(0,y)e^{-iHt}
    \end{align}
    \item {\it Dressed} operators: Solutions to the flow equation
    \begin{align}\label{flowedeq}
        \mathcal{D}_\lambda \tilde{\mathcal{O}} = 0, \quad \tilde{\mathcal{O}}|_{\lambda=0} = \mathcal{O}^{(0)}
    \end{align}
\end{itemize}
If we take a basis of local operators of the seed CFT at $t=0$, the definition \eqref{udeq} evolves it to an arbitrary spacetime point via Hamiltonian evolution \footnote{In this paper, by local operators we refer to operators that can be expressed in terms of fundamental fields at a given point on the constant time slice, without any integrals.  Non-local operators refer to operators involving the integral of local operators. In this manuscript, this  definition is used only in the classical sense. If we start with a basis of operators which commute on the initial time slice, the undeformed operators remain commuting with each other on an arbitrary time slice. }. However, undeformed operators lack clear organization under the {Virasoro$\times$Virasoro symmetries}{}. Dressed operators, in contrast, preserve CFT$_2$ symmetry properties, including Ward identities and correlation function structures, but are explicitly non-local. This motivates our search for a third operator class that combines locality with symmetry organization.

We introduce {\it physical} operators in $T\bar{T}$-deformed theories:
\begin{itemize}
    \item {\it Physical} operators: At the classical level, physical operators can be defined by a local expression in terms of the undeformed operators and stress tensor,
    \begin{equation}\label{physicalOp}
        \mathcal{O}_{h,\bar{h}} \equiv \frac{1 + \lambda^2 \mathcal{O}_{T\bar{T}}}{\left(1 + 2\lambda \mathcal{H}_R\right)^h \left(1 + 2\lambda \mathcal{H}_L\right)^{\bar{h}}} \mathcal{O}_{h,\bar{h}}^{(0)}
    \end{equation}
\end{itemize}
{where} ${\MO}_{h,\bar{h}}^{(0)}$ is an undeformed operator whose pre-image in the CFT is a primary operator with conformal weights $(h,\bar h)$, $\mathcal{O}_{T\bar{T}}$ is the $T\bar T$ operator, and $\mathcal{H}_{L/R}$ are stress tensor components. 
The physical operators admit a reformulation in terms of dressed operators:
\begin{align}\label{Ophysintro}
    &\mathcal{O}_h(u,v)  = \int d\sigma^2 \, \delta(U-u, V-v) \, \tilde{\mathcal{O}}_{h,\bar{h}}(\sigma^+,\sigma^-) 
\end{align}
Here $\tilde{\mathcal{O}}_{h,\bar{h}}$ denotes a dressed primary operator, $u,v$ are null coordinates, and $U$ and $V$ are nonlocal coordinates that depend on the stress tensor. We will show
that the physical operator is equivalent to the operator considered in \cite{Aharony:2023dod} at the classical level.
This reformulation \eqref{Ophysintro} of physical operators provides a symmetry-guided framework that systematically simplifies the computation of physical quantities such as correlation functions. 

At the quantum level, we assume that the dressed operators still exist and transform under the dressed conformal symmetry as in a CFT. Using the dressed operators and dressed stress tensor as building blocks, we define the quantum version of the physical operators by promoting the relation \eqref{Ophysintro}, and its momentum space expression to the quantum level. 
We compute the two-point correlation function of physical operators in momentum space, summing the most UV-sensitive terms to obtain the non-perturbative result:
\begin{align}\label{Gnintr}
    \langle \mathcal{O}_{h,\bar{h}}(p,\bar{p}) \mathcal{O}_{h,\bar{h}}(-p,-\bar{p}) \rangle &\sim 
    \frac{\pi \Gamma(1-2h_\lambda)}{\Gamma(2h_\lambda)} \left(\frac{|p|}{2}\right)^{4h_\lambda-2}
\end{align}
This exactly matches both the string theory result via TsT/$T\bar{T}$ correspondence~\cite{Cui:2023jrb} and field theory calculations in the large momentum limit~\cite{Aharony:2023dod}. 

The remainder of this paper is organized as follows. In Sec.\ref{sec: reviews}, we review the Hamiltonian formalism of $T\bar T$ deformed CFTs, their relation with canonical transformation, and then discuss symmetries and operators. 
In Sec.3, we explicitly construct the {dressed} stress tensor and nonlocal symmetry generators on the plane, by solving the flow equation \eqref{flowedeq} for the stress tensor. 
In Sec.4, we construct the dressed operators by solving the flow equation \eqref{flowedeq} for primary operators.
In Sec.5, we introduce
a classical definition of physical operators
and discuss their relationship to alternative operator formulations. In Sec.6, we compute physical operator correlation functions via systematic series expansion, and obtain a non-perturbative result by summing over the most UV-sensitive terms.

\section{$T\bar T/J\bar T$ deformation as a canonical transformation} \label{sec: reviews}

$T\bar T$ deformations can be defined in any quantum field theory in two dimensions with a conserved stress tensor \cite{Zamolodchikov:2004ce,Smirnov:2016lqw,Cavaglia:2016oda}, and can also be reformulated in many useful ways, such as using flat JT gravity \cite{Dubovsky:2017cnj,Dubovsky:2018bmo}, random geometry \cite{Cardy:2018sdv}, effective string theory \cite{Cavaglia:2016oda,McGough:2016lol,Callebaut:2019omt}, and Hamiltonian formalism \cite{Jorjadze:2020ili,Kruthoff:2020hsi}. 
$T\bar{T}$ deformations can also be generalized to theories without Lorentzian symmetry \cite{Esper:2021hfq}, and to dimensions other than two \cite{Hartman:2018tkw,Taylor:2018xcy,Gross:2019uxi}. 
Similar properties are also shared by other irrelevant deformations, including $J\bar T$ \cite{Guica:2017lia,Chakraborty:2018vja,Apolo:2018qpq}, and combinations of $T\bar T$ and $J\bar T$ deformations \cite{Kutasov:2019mdf,LeFloch:2019rut,Frolov:2019xzi,Apolo:2021wcn}.
    
 In this section, we establish conventions and review a class of solvable deformations in two-dimensional quantum field theories, such as $T\bar T$ and $J\bar T$
  deformations, focusing on the Hamiltonian formalism developed in \cite{Jorjadze:2020ili,Kruthoff:2020hsi, Georgescu:2024ppd}. In what follows, we aim to discuss these deformations uniformly. 

  For any two-dimensional quantum field theory equipped with two commuting global symmetry currents, a class of current-current type deformations is constructed by incrementally adding a composite operator to the action:
\begin{equation}\label{Sd}
    {\partial S\over \partial \lambda}= \int dx^2  \, \mathcal{O}_{d.f} ,\quad \mathcal{O}_{d.f}= - {1\over 2}\epsilon^{ab} \epsilon^{\mu\nu}  J_{\mu a } J_{\nu b}
\end{equation}
where $\lambda$ is the deformation parameter, and $S$ is the deformed action at $\lambda$. The Noether currents $J^{(a)}$ satisfy the conservation law $ d*J^{(a)}=0$. Different choices of the currents correspond to different types of deformations. For example, if we consider translational symmetries along the light-cone coordinates, we obtain the $T\bar T$ deformation with \begin{align}
  J_{\underline{u}}&=T_{uu} du + T_{u v}d v,\qquad
 J_{\underline{v}} = T_{vu} du + T_{vv}d v, 
\end{align}
where $u = y + t, v = y -t$ are  the lightcone coordinates. 
In this context, the deformation operator is defined as 
\begin{equation} 
\mathcal{O}_{T\bar{T}} = -\det\left(T_{\mu}^{\ \nu}\right) 
\end{equation}
Although its name suggests otherwise, this operator is generally not proportional to the product of the left and right-moving stress tensors, except in the case of infinitesimal deformations about conformal field theories.

Similarly, for $J\bar{T}$ deformation, the currents are:
\begin{align}
  J_{\underline{1}}&=J_{u} du + J_{v}d v,\qquad
 J_{\underline{2}} = T_{vu} du + T_{vv}d v
\end{align}    

    \subsection{Undeformed and {dressed} operators}

The current-current-type deformation can also be discussed in the Hamiltonian formalism \cite{Kruthoff:2020hsi} \footnote{At the classical level, the Lagrangian and Hamiltonian formalisms are equivalent, while subtleties related to operator ordering appear at the quantum level. }. On the equal time slice, the deformed Hamiltonian can be defined as 
\begin{align}\label{dH}
      \p_\lambda H&=-\int dy\, \MO_{d.f}=\frac{1}{2}\int dy dy'\epsilon^{\mu\nu}\epsilon^{ab}J_{\mu a}(y)J_{\nu b}(y')\delta(y-y')
    \end{align}
    where we use $a, b$ to specify different Noether currents, and $\mu,\nu$ specify the spacetime components. 
   To reduce clutter, we use quantities without extra superscripts for the deformed theory throughout this paper.
The main idea of Hamiltonian formalism is to reformulate the deformation \eqref{dH} as an evolution with respect to the deformation parameter. In order to do so, let us introduce a Green's function on the spatial slice with period $R
$, \begin{align}
    \p_y G(y-y')=\delta(y-y')-{1\over R}
\end{align}
Replacing the delta function by the Green's function defined above and using the conservation laws of the Noether currents, the deformation of the Hamiltonian can be written as a sum of two terms 
\begin{align}\label{dHxy}
      \p_\lambda H&=i[{\mathcal{X}},H]+\mathcal{Y}
    \end{align}
where
   \begin{align}
&{\mathcal{X}}=\frac{1}{2}\int dy dy'G(y-y')J_{ta}(y)J_{tb}(y')\epsilon^{ab}\\
&\mathcal{Y}=\frac{1}{2R}\int dy dy'\epsilon^{\mu\nu}\epsilon^{ab}J_{\mu a}(y)J_{\nu b} (y')
\end{align} 
 The above decomposition allows us to discuss the flow of states and operators in a simple way.  The first term in \eqref{dHxy} corresponds to a canonical transformation on the phase space which does not change the spectrum, and the second term $\mathcal Y$ is a sum of factorized operators which is responsible for the change of eigenvalues on the cylinder. In the following, we will focus on the theory on the plane with $R\to\infty$, so that $\mathcal Y=0$ and the deformation can be regarded as a canonical transformation which mixes different states in the Hilbert space without changing the energy spectrum.

 In this paper, we will consider three types of operators in the deformed theory: {\it undeformed } operators, {\it {dressed} } operators, and {\it physical } operators. The first two operators have been discussed previously, and we briefly review them below. We will propose a third type of operators, the {\it physical } operators, in section 5.
 
 The {\it undeformed } operators are defined by evolving the undeformed states on an initial values slice by the deformed Hamiltonian, namely
 \begin{equation} \label{undeformedO}
        \MO^{(0)}(t,y)\equiv e^{iHt}\MO^{(0)}(0,y)e^{-iHt}
    \end{equation}
The undeformed operators are the natural operators to define in the deformed theory.
These operators will no longer be organized by symmetries along the flow, which makes the computation of physical quantities such as the correlation functions and structure constants very challenging. 

The $dressed$ operators \cite{Kruthoff:2020hsi}, also called the $flowed$ operators in \cite{Guica:2020eab,Guica:2021fkv},  are motivated by the observation that the flow equation of the Hamiltonian \eqref{dHxy} 
corresponds to an evolution along $\lambda$, generated by the operator $\mathcal X$. This can be phrased geometrically in the space of QFTs with coordinate $\lambda$ and connection $\mathcal X$. Then the $T\bar{T}$ deformation implements parallel transport of the Hamiltonian via the covariant derivative, 
\begin{align}\label{Hflow}
\mathcal{D}_\lambda H=0  ,\quad  \mathcal{D}_\lambda \equiv \p_\lambda-i[{\mathcal{X}},\cdot] 
\end{align}
The dressed operators are defined by requiring that they are covariant constants under the $T\bar T$ flow,  
    \begin{equation}\label{fO}
        \mathcal{D}_\lambda \tilde{\mathcal{O}}=0,\quad \tilde{\mathcal{O}}|_{\lambda=0}=\mathcal{O}^{(0)}
    \end{equation}
    where the last equation is the initial condition.

    Dressed operators preserve several nice properties along the flow, including chirality, conservation laws, and the algebra between {dressed} operators themselves \cite{Kruthoff:2020hsi,Georgescu:2024ppd}. The reason is that derivatives with respect to the deformation parameter $\lambda$ and coordinates can be exchanged, and that the Hamiltonian satisfies \eqref{Hflow}.

   For instance, the flow of the chirality condition satisfies,  
       \begin{align}
 \p_\lambda  \z({\breve\p_t}\tilde{\MO}+i[H,\tilde{\MO}]-\p_y\tilde{\MO}\y)=i \z[{\mathcal{X}}, \z({\breve\p_t}\tilde{\MO}+i[H,\tilde{\MO}]-\p_y\tilde{\MO}\y)\y],   \label{chiralflow}
    \end{align}
     where ${\breve\p_t}$ denotes derivative with respect to explicit dependence of time. 
    For a {dressed} operator built from a chiral operator in the undeformed CFT, the chirality condition is initially satisfied 
    \begin{align}   \quad \z({\breve\p_t}\tilde{\MO}+i[H,\tilde{\MO}]-\p_y\tilde{\MO}\y) \Big|_{\lambda=0}=0\label{chirali} \end{align}
and then the flow equation \eqref{chiralflow} will make sure that the operator remains chiral along the deformation flow.

Similarly, if the undeformed theory has a conserved current 
    $ J^{(0)}_\mu$, the  $flowed$ current $\tilde{J}_\alpha$ defined by the above condition \eqref{fO} remains conserved under the deformation. This can be seen directly from the flow equation
    \begin{equation}  \label{conserved} \p_\lambda\z({\breve\p_t}\tilde{J}_t+i[H,\tilde{J}_t]-\p_y\tilde{J}_y\y)=i \z[{\mathcal{X}},\z(\breve\p_t\tilde{J}_t+i[H,\tilde{J}_t]-\p_y\tilde{J}_y\y)\y]
    \end{equation}
which again preserves the conservation law along the flow.

Furthermore, the covariant constant condition \eqref{fO} also preserves Ward identities, so that {dressed} operators transform under the {dressed} symmetry generators just like in the undeformed theory. When the undeformed theory is a CFT, we expect to be able to find the {dressed} version of Virasoro generators and primary operators. The {dressed} operators satisfy nice transformation rules under the symmetries, and their correlation functions and structure constants remain unchanged under the flow. However, they are related to the undeformed ones in a non-local way, which makes their physical interpretation obscure.

To summarize, the undeformed operators are local but are not organized by symmetry in a manageable way. The {dressed} operators have Ward identities and correlation functions as in a CFT$_2$,  but are nonlocal. As we will show in section 5, 
the newly proposed physical operators show up as a bridge between the aforementioned two types of operators, and have both advantages: classically local and symmetry-organized.

\section{Nonlocal conformal symmetries in $T\bar{T}$ deformation}\label{TTbarsymmetry}

In this section, we discuss classical symmetries in the $T\bar{T}$-deformed  CFT$_2$ on the plane.  From the definition of the {dressed} operators and {dressed} currents, we learn that the symmetries in the undeformed theory are preserved by the deformation, albeit realized in a non-local way. The dressed symmetries generators of the $T\bar{T}$ deformed CFT$_2$ on the cylinder have been derived in \cite{Guica:2022gts}, which after taking the large radius limit reduces to those we obtain directly on the plane. In the following, we will first express the deformed stress tensor in terms of the undeformed one using Noether’s theorem within the Hamiltonian formalism, and then construct the {dressed} stress tensor by solving the covariant constant condition given in \eqref{fO}.

Let us schematically denote the action of the $T\bar T$ deformed CFT$_2$ by 
    \begin{equation}
        S[\phi]=\int d^{2} x \, \mathcal{L}(\phi(x), \partial_{\mu} \phi(x))
    \end{equation}
    where $\phi$ collectively denotes the fundamental fields. 
   The $T\bar{T}$ deformation preserves translational invariance, which enables us to obtain the stress tensor from Noether's theorem, \begin{equation}  \label {Neother}T_\mu\,^\nu=\eta_\mu\,^\nu\mathcal L- \partial_\mu\phi{\partial \mathcal{L}\over  \partial{\partial _\nu \phi}}
   \end{equation}
   By a Legendre transformation, we have  the Hamiltonian density  \begin{equation}  
\mathcal{H} = \pi \partial_{t} \phi - \mathcal{L},\quad \pi={\partial\mathcal{L}\over \partial\partial_t \phi}
  \end{equation} 
so that the stress tensor\footnote{Note that the stress tensor \eqref{Tuv} does not apply to theories like Liouville theory, which has an improved term in the stress tensor.} can be written as
    \begin{equation}
        \begin{aligned}   \label{Tuv}
        & T_{t t}=\mathcal{H}, \quad T_{t y}=\frac{\partial \mathcal{H}}{\partial\left(\partial_y \phi\right)} \frac{\partial \mathcal{H}}{\partial \pi}=T_{y t}=\pi \partial_y \phi\equiv \mathcal{P} \\
        & T_{y y}= \hat {V}[ \mathcal{H}]-\mathcal{H}, \quad \hat {V}=\pi \frac{\partial}{\partial \pi}+\partial_y \phi\frac{\partial }{\partial\left(\partial_y \phi\right)} 
        \end{aligned}
    \end{equation}
    where we have used the equations of motion, and assumed that we have already adjusted the theory so that the stress tensor is already symmetric.    As is shown in \cite{Jorjadze:2020ili}, this condition is preserved along the $T\bar{T}$ deformation. 

Under the  $T\bar T$ deformation\begin{equation}\label{flowH}
\partial_\lambda \MH =-
\mathcal{O}_{T\bar{T}} = \det\left(T_{\mu}^{\ \nu}\right)=\MP^2-\MH T_{yy}
\end{equation}
Now let us solve the flow equation at $t=0$. The deformed energy density can be regarded as a functional of $\MP$ and the deformed energy density $\MH^{(0)}$, so that
\begin{equation}
    T_{yy}=\hat V[\MH^{(0)}] {\partial \MH\over \partial \MH^{(0)}}+\hat V[\MP] {\partial \MH\over \partial \MP}=2 \MH^{(0)} {\partial \MH\over \partial \MH^{(0)}}+2\MP {\partial \MH\over \partial \MP}\label{Tyy}
\end{equation}
where we have used the initial condition that the undeformed theory is a CFT$_2$ with traceless condition $\hat V[\MH^{(0)}]=2 \MH^{(0)}$. 
Then the flow equation \eqref{flowH}  can be solved by:
    \begin{equation} \label{deformedH}
        \mathcal{H}=\frac{1}{2 \lambda }\left(\sqrt{1+4 \lambda \mathcal{H}^{(0)}+4 \lambda^2 \mathcal{P}^2}-1\right)
    \end{equation}
It is convenient to introduce energy density in the lightcone coordinates.
\begin{equation}
    \mathcal{H}_{L} \equiv T_{ut}=\frac{\mathcal{H} + \mathcal{P}}{2}, \quad \mathcal{H}_{R} \equiv -T_{vt}= \frac{\mathcal{H}- \mathcal{P}}{2}, \quad u=y+t,\quad v=y-t
\end{equation}
The relation between the deformed and undeformed stress tensor can be reversed so that
\begin{equation}\label{RST}
	\MH_L^{(0)}=(1+2\lambda\MH_R)\MH_L,\quad \MH_R^{(0)}=(1+2\lambda\MH_L)\MH_R
\end{equation} 
Plugging the solution \eqref{RST} into \eqref{Tyy},
we obtain \begin{align}
    T_{y y}=  \frac{\mathcal{H} + 2 \lambda \mathcal{P}^{2}}{1+ 2 \lambda \mathcal{H}}\label{Tyys}
\end{align} and the deformation operator can be written as
\begin{equation} \label{Ottbar}
        \OT 
        = \frac{4\mathcal{H}_{L} \mathcal{H}_{R}}{1+2 \lambda \mathcal{H}} 
    \end{equation}
Although the relation \eqref{RST} between deformed and undeformed stress tensor is obtained by solving the flow equation at $t=0$, it can be evolved to arbitrary time by the deformed Hamiltonian. 
In other words, the relation \eqref{RST} is valid at arbitrary constant-time slices, provided that the operators at $t$ are related to those at $t=0$ by  
\begin{align}\label{RSTt}
\MH_{L/R}^{(0)}(t,y)&=e^{iHt}\MH_{L/R}^{(0)}(0,y)e^{-iHt}, \quad \MH_{L/R}(t,y)=e^{iHt}\MH_{L/R}(0,y)e^{-iHt}
\end{align} 
Note that the above definition of the undeformed stress tensor is consistent with the notion of the undeformed operator  \eqref{undeformedO}. Despite the name, undeformed operators only agree with operators in the undeformed $\text{CFT}_2$ at $t=0$. At a later time, the undeformed operator $\mathcal {O}^{(0)}$ in the $T\bar T$ deformed theory is evolved by the deformed Hamiltonian, whereas operators in the $CFT_2$ are evolved by the undeformed one.

\subsection{The {dressed} stress tensor}
In order to find the {dressed} stress tensor in $T\bar T$ deformed CFT$_2$, we need to solve the flow equation \eqref{fO}, which at the classical level is given by 
\begin{align}\label{feqT}
\MD_\lambda\tilde{\MH}_L=\partial_\lambda\tilde{\MH}_L+\{ \mathcal{X}_{T\bar{T}}, \tilde{\MH}_L\}=0, \quad \MD_\lambda\tilde{\MH}_R=0
\end{align}
where the flow generator 
 for $T\bar T$ deformation is given by
 \begin{align}
{\mathcal{X}}_{T\bar{T}}&=\frac{1}{2}\int dy dy'G(y-y')\z(\MH(y)\MP(y')-\MP(y)\MH(y')\y)
\\&=\int dy dy'G(y-y')\z(\MH_R(y)\MH_L(y')-\MH_L(y)\MH_R(y')\y)\nonumber
    \end{align}
To solve the covariant constant condition \eqref{feqT}, it is useful to use the expressions of
Poisson brackets among the components of the stress tensor \cite{Guica:2022gts} \footnote{The Poisson brackets here do not have a classical central term, which is a consequence of the choice of the stress tensor \eqref{Tuv}. It is interesting to explicitly work out the cases where a classical central appears, which we leave for future study. }, 
\begin{equation}
\begin{aligned}\label{HP}
& \left\{\mathcal{H}(y), \mathcal{H}\left(y^{\prime}\right)\right\}=\left(\mathcal{P}(y)+\mathcal{P}\left(y^{\prime}\right)\right) \partial_y \delta\left(y-y^{\prime}\right) \\
& \left\{\mathcal{P}(y), \mathcal{P}\left(y^{\prime}\right)\right\}=\left(\mathcal{P}(y)+\mathcal{P}\left(y^{\prime}\right)\right) \partial_y \delta\left(y-y^{\prime}\right) \\
& \left\{\mathcal{H}(y), \mathcal{P}\left(y^{\prime}\right)\right\}=\left(T_{y y}(y)+\mathcal{H}\left(y^{\prime}\right)\right) \partial_y \delta\left(y-y^{\prime}\right)
\end{aligned}
\end{equation}
Plugging in the explicit expression \eqref{Tyys}, we find
\begin{equation}\label{HLRp}
	\begin{aligned}
	&\{\mathcal{H}_L(y_1),\mathcal{H}_L(y_2)\}= \bigg[\p_{y_1}\z(\frac{\lambda}{2}\OT(y_1)-\mathcal{H}_L(y_1)\y)+2\z(\HL(y_2)-\frac{\lambda}{2}\OT(y_2)\y)\p_{y_1}\bigg] \delta(y_1-y_2)\\
	&\{\mathcal{H}_R(y_1),\mathcal{H}_R(y_2)\}= \bigg[\p_{y_1}\z(\HR(y_1)-\frac{\lambda}{2}\OT(y_1)\y) -2\z(\HR(y_2)-\frac{\lambda}{2}\OT(y_2)\y)\p_{y_1}\bigg] \delta(y_1-y_2)\\
	&\{\mathcal{H}_L(y_1),\mathcal{H}_R(y_2)\}=-\frac{\lambda}{2}\p_{y_1}\mathcal{O}_{T\bar{T}}(y_1)\delta(y_1-y_2)
	\end{aligned}
\end{equation}  
At the CFT point $\lambda=0$, the above Poisson brackets imply that $\MH_L$ and $\MH_R$ are the left and right moving stress tensors, but without a central term. 
Using the Poisson brackets \eqref{HLRp}, one finds 
\begin{equation}
\mathcal{D}_\lambda \mathcal{H}_{L}=\partial_y \Big(\mathcal{D}_\lambda \eta_0\Big), \quad \mathcal{D}_\lambda \mathcal{H}_{R}=\partial_y \Big(\mathcal{D}_\lambda {\bar \eta}_0\Big)
\end{equation}
where
\begin{equation}
	\begin{aligned}
		&\eta_0=\int dy'G(y-y') \mathcal{H}_{L}(y'),\quad {\bar \eta}_0=\int dy'G(y-y') \mathcal{H}_{R}(y').
	\end{aligned}
\end{equation}
and 
\begin{equation}
    \mathcal{D}_\lambda \eta_0 = 2 \, \bar\eta_0 \, \mathcal{H}_{L} + \lambda  \z(\eta_0-{\bar\eta}_0\y) \,\mathcal{O}_{T \bar{T}},\quad \mathcal{D}_\lambda {\bar\eta}_0 = 2 \, \eta_0 \, \mathcal{H}_{R} + \lambda \z({\bar \eta}_0-{\eta}_0\y) \,\mathcal{O}_{T \bar{T}}
\end{equation}
Using induction, one can verify the following useful relation 
\begin{equation}\label{TTbarsymmetryinduction}
    \mathcal{D}_\lambda\z(\frac{1}{k!}\HL(-2\lambda {\bar\eta}_0)^k\y)=\p_{y}\z(\frac{1}{k!}\mathcal{D}_\lambda \eta_0(-
   2\lambda \bar \eta_0)^k\y)-\frac{1}{(k-1)!}(-2\lambda {\bar\eta}_0)^{k-1}\mathcal{D}_\lambda \eta_0
\end{equation}
The above relation enables us to construct the solution of the flow equation \eqref{feqT}  
\begin{align}	\label{flowed HLR}\tilde{\mathcal{H}}_L&\equiv\sum_{n=0}^\infty \frac{(-1)^n}{n!}\p_y^n \left( \mathcal{H}_L\delta \hat u^n\right) ,\quad \tilde{\mathcal{H}}_{R}\equiv\sum_{n=0}^\infty \frac{(-1)^n}{n!}\p_y^n \left( \mathcal{H}_{R}\delta\hat v^n\right)\\
\delta \hat u&=2\lambda {\bar\eta}_0, \quad  \delta \hat{v} = 2 \lambda \eta_{0}, \nonumber
\end{align}
which satisfy 
\begin{equation}\label{HLDcd}
\MD_\lambda\tilde{\mathcal{H}}_L=\MD_\lambda\tilde{\mathcal{H}}_R=0
\end{equation}
By using the {dressed} stress tensor, we can construct an infinite number of symmetry generators 
\begin{equation} \label{nonlocalL}
	\begin{aligned}
	\tilde{L}_m &=i^{m+1}\int dy \, u^{m+1} \tilde{\mathcal{H}}_L 
		=i^{m+1}\int dy\,(u+\delta \hat u)^{m+1}\mathcal{H}_L=i^{m+1}\int dy\,{\hat u}^{m+1}\mathcal{H}_L
	\end{aligned}
\end{equation}
where we have used integration by parts in the second equality and introduced a nonlocal coordinate 
\begin{equation}\label{nonlocal0}
    \begin{aligned}
        &\hat{u}\equiv u+\delta \hat{u},\quad \hat{v}\equiv v +\delta\hat{v},\quad \delta \hat{u}=2\lambda {\bar\eta}_0,\quad \delta\hat{v}=2\lambda \eta_0
    \end{aligned}
\end{equation}
in the last equality. Expression \eqref{nonlocalL} for the charge is formal. For $m>1$, the charges become finite for local excitations with sufficiently rapid fall-off at infinity.
Then the {dressed} symmetry generators flow covariantly along the deformation due to \eqref{HLDcd}, and are hence conserved as the $T\bar T$ flow preserves the conservation law  \eqref{conserved}. We have 
\begin{equation}
  \mathcal{D}_\lambda \tilde{L}_m=0, \quad  \p_t \tilde{L}_m={\breve\p_t}\tilde{L}_m+i[H,\tilde{L}_m]=0
\end{equation}
Similarly, we have the anti-holomorphic part of the symmetries
\begin{equation}\label{barLm}
    \tilde{\bar{L}}_m=(-i)^{m+1}\int dy(y-t+2\lambda \eta_0)^{m+1}\HR
\end{equation}
As expected, the {dressed} symmetry generators form the same algebra as the undeformed ones, namely
the left and right moving Witt algebra,
\begin{equation}\label{Witt}
    \begin{aligned}
        &i\{\tilde{L}_n,\tilde{L}_m\}=(n-m)\tilde{L}_{n+m}, \quad
        i\{\tilde{\bar{L}}_n,\tilde{\bar{L}}_m\}=(n-m)\tilde{\bar{L}}_{n+m}, \quad
        i\{\tilde{L}_n,\tilde{\bar{L}}_m\}=0
    \end{aligned}
\end{equation}
The symmetry algebra obtained so far does not have a central term. In a CFT$_2$, the central term can have two origins: classical central charge and quantum central charge. The construction so far has focused on theories without a classical central charge. The dressed symmetry generators \eqref{nonlocalL}, \eqref{barLm} and their algebra are compatible with the large radius limit of the result on the cylinder \cite{Guica:2022gts}. 

We have argued that the chirality condition will be preserved along the flow \eqref{chiralflow}. 
Now let us check this directly with the explicit form of the dressed stress tensor \eqref{flowed HLR}. The full dependence on a coordinate includes both the explicit and the implicit parts, so that the derivative can be written as   
\begin{equation}
\partial_u\equiv\frac{1}{2}\z(\p_y+{\breve{\p}}_t+\left\{\, \cdot \, ,H\right\}\y) , \quad \partial_v\equiv\frac{1}{2}\z(\p_y-{\breve{\p}}_t-\left\{ \, \cdot \, ,H\right\}\y) \label{du}
\end{equation}
where ${\breve{\p}}$ denotes taking derivative with respect to the explicit time dependence, and $\p_y$ is the full derivative with respect to $y$.
Note that acting on $\eta_0, \bar\eta_0$ and all local operators, $\p_y$  is equivalent to $\{\cdot, P\}$, and $\p_u$  is equivalent to $\{\cdot, H_L\}$.
Then we have
 \begin{align}
\partial_v \tilde{\MH}_L=\{\tilde{\MH}_L,H_R\}=\{ \tilde{\MH}_L,{\tilde{\bar L}}_{-1}\}=0,\quad  
\partial_u \tilde{\MH}_R=-\{\tilde{\MH}_R,H_L\}=\{ \tilde{\MH}_R, {\tilde{L}}_{-1}\}=0
 \end{align}
 which means that the {dressed} stress tensors are holomorphic and anti-holomorphic.

 From the expression \eqref{nonlocalL} and \eqref{barLm}, we note that the Lorentzian boost generator is given by
 \begin{align}\label{boost}
   M\equiv i\int dy (u\MH_L+v \MH_R)=i\int dy (u\tilde{\MH}_L+v \tilde{\MH}_R)= \tilde{L}_0-\tilde{\bar{L}}_0
 \end{align} which is also a dressed operator. Within the nonlocal dressed conformal symmetry generators, we find a subalgebra generated by translations $\tilde{L}_{-1}=H_{L},\,\bar{\tilde{L}}_{-1}=H_{R}$ and Lorentzian boost $M=\tilde{L}_0-\tilde{\bar{L}}_0$. This is compatible with the general expectation that $T\bar T$ deformation preserves Lorentzian invariance.

As a final remark, we note that the deformed stress tensor has the following Poisson brackets with the {dressed} symmetry generators, 
\begin{equation}
   \begin{aligned}
   i\{\tilde{L}_m,\HL\}&=i^m(m+1)\Big(2+2\lambda^2 \OT\Big)(u+\delta\hat{u})^m\HL+i^m(u+\delta\hat{u})^{m+1}\p_u\HL\\
   &+2\lambda (m+1)\eta_m\p_v\HL\\
   i\{\tilde{\bar{{L}}}_m,\HL\}&=(-i)^m(m+1)\Big(\frac{4\lambda\HR}{1+2\lambda\MH}+2\lambda^2 \OT\Big)(v+\delta\hat{v})^m\HL+(-i)^m(v+\delta\hat{v})^{m+1}\p_v\HL\\
   &+2\lambda (m+1)\bar{\eta}_m\p_u\HL+(-i)^m(m+1)(v+\delta\hat{v})^{m}\lambda\OT\\
   i\{\tilde{L}_m,\HR\}&=i^m(m+1)\Big(\frac{4\lambda\HL}{1+2\lambda\MH}+2\lambda^2 \OT\Big)(u+\delta\hat{u})^m\HR+i^m(u+\delta\hat{u})^{m+1}\p_u\HR\\
   &+2\lambda (m+1)\eta_m\p_v\HR+i^m(m+1)(u+\delta\hat{u})^{m}\lambda\OT\\
   i\{\tilde{\bar{{L}}}_m,\HR\}&=(-i)^m(m+1)\Big(2+2\lambda^2 \OT\Big)(v+\delta\hat{v})^m\HR+(-i)^m(v+\delta\hat{v})^{m+1}\p_v\HR\\
   &+2\lambda (m+1)\bar{\eta}_m\p_u\HR
\end{aligned}
\end{equation}
where we have defined
\begin{equation}
\begin{aligned}\label{etam}
   &	\eta_m(t,y)\equiv i^m\int dy'G(y-y')(y'+t+\delta\hat{u})^m\HL(y')\\
   &\bar{\eta}_m(t,y)\equiv (-i)^m\int dy'G(y-y')(y'-t+\delta\hat{v})^m\HR(y')
\end{aligned}
\end{equation}

\subsection{The non-local coordinates}
From the construction of symmetry generators, we can see the non-local coordinates emerge \eqref{nonlocal0}, which we reproduce here for convenience
\begin{equation}\label{nonlocal}
    \begin{aligned}
        &\hat{u}\equiv u+\delta \hat{u},\quad \hat{v}\equiv v +\delta\hat{v},\quad \delta \hat{u}=2\lambda {\bar\eta}_0,\quad \delta\hat{v}=2\lambda \eta_0
    \end{aligned}
\end{equation}
From the definition, we can calculate
\begin{equation}
    \begin{aligned}
        &\p_u\hat{u}=1+\lambda^2\OT,\quad \p_v\hat{u}=2\lambda\HR-\lambda^2\OT\\
        &\p_u\hat{v}=2\lambda\HL-\lambda^2\OT,\quad \p_v\hat{v}=1+\lambda^2\OT
    \end{aligned}
\end{equation}
We can invert the relation and obtain 
\begin{equation}\label{Inverse}
    \begin{aligned}
        &\p_{\hat{u}} u=1,\quad \p_{\hat{v}} u=-\frac{2\lambda \HR(u,v)}{1+2\lambda\HL(u,v)} \\
        &\p_{\hat{u}} v=-\frac{2\lambda \HL(u,v)}{1+2\lambda\HR(u,v)},\quad \p_{\hat{v}} v=1
    \end{aligned}
\end{equation}
which can be integrated so that 
\begin{align}\label{inverse}
    u&=\hat u-2\lambda \int d\hat v' G(\hat v-\hat v')\tilde \MH_R(\hat v') =\hat u-2\lambda \int d v' G(\hat v- v')\tilde \MH_R( v')\\
    v&=\hat v-2\lambda \int d\hat u' G(\hat u-\hat u')\tilde \MH_L(\hat u') =\hat v-2\lambda \int d u' G(\hat u- u')\tilde \MH_L( u')
\end{align}
where in the last step, we have renamed $\hat v'$ by $v'$ as it is a dummy variable. 

From the expression of the {dressed} stress tensor \eqref{flowed HLR}, we obtain the following useful relation by using integration by parts
\begin{equation}\label{fH}
	\begin{aligned}
	&\int dy' f(y')\tilde{\mathcal{H}}_L(t+y')=\int dy' f(y'+\delta\hat{u}(t,y'))\HL(t,y')\\
    &\int dy'f(y')\tilde{\mathcal{H}}_R(y'-t)=\int dy' f(y'+\delta\hat{v}(t,y'))\HR(t,y')
	\end{aligned}
\end{equation}
where all quantities above are assumed to be on the constant time slice. 
By assuming the integral is evaluated on configurations that vanish sufficiently fast at infinity, we have dropped the total derivative terms. Some interesting choices of $f(y)$ are as follows. 
\begin{itemize}
    \item 
 Choosing $f(y')=\delta(y'-(y+\delta\hat{u}))$ for $\HL$ and $f(y')=\delta(y'-(y+\delta\hat{v}))$
for $\HR$ in \eqref{fH}, we find an equal time relation between the {dressed} stress tensor and the physical one, 
\begin{align}\label{Hrelation}
	\begin{aligned}
		&\tilde{\mathcal{H}}_L(\hat{u})=\frac{\HL(t,y)}{1+2\lambda\HR(t,y)},
		\quad \tilde{\mathcal{H}}_R(\hat{v})=\frac{\HR(t,y)}{1+2\lambda\HL(t,y)}
	\end{aligned}
\end{align}
where  $y+t+\delta u$ is replaced by $\hat u$.
This is the relation derived from the Lagrangian formulation of the $T\bar{T}$ deformation \cite{Guica:2020uhm} and the holographic side \cite{Guica:2019nzm, Kraus:2021cwf}. Now we have derived this result from the Hamiltonian formalism.
Combining with the relation between the physical stress tensor and the undeformed stress tensor \eqref{RST}, we obtain another relation between the  {dressed} stress tensor and the undeformed stress tensor, 
\begin{align}\label{H0relation}
	\begin{aligned}
		&\tilde{\mathcal{H}}_L(\hat{u})=\frac{\HL^{(0)}(t,y)}{(1+2\lambda\HR(t,y))^2}=({\partial \hat u\over \partial y})^{-2} \HL^{(0)}(t,y),\\
		&\tilde{\mathcal{H}}_R(\hat{v})=\frac{\HR^{(0)}(t,y)}{(1+2\lambda\HL(t,y))^2}=({\partial \hat v\over \partial y})^{-2} \HR^{(0)}(t,y)
	\end{aligned}
\end{align}

\item By choosing $f(y')=\delta(y-y')$, we can express the dressed stress tensor in terms of the physical stress tensor by an integral
\begin{align}
    \tilde{\MH}_L(u)=\int dy'\delta(y-y'-\delta \hat u')\MH_L(t,y')=\int dy'\delta(u-\hat u')\MH_L(t,y')
\end{align}
where $\delta \hat u'=\delta \hat u(t,y')$, and $ \hat u'=t+\delta \hat u(t,y')$. 
\item 
By choosing $f(y')=G(y+\delta\hat{u}-y')$ for $\HL$ and $f(y')=G(y+\delta\hat{v}-y')$ for $\HR$, we have 
\begin{equation}\label{Ginverse}
    \begin{aligned}
        &\int dy' G(y+\delta\hat{v}(t,y)-y')\tilde{\MH}_R(y'-t)=\eta_0(t,y)\\
        &\int dy' G(y+\delta\hat{u}(t,y)-y')\tilde{\MH}_L(y'+t)={\bar\eta}_0(t,y)
    \end{aligned}
\end{equation}
This provides another derivation of the inverse relation between the coordinates and non-local coordinates.
\begin{equation}\label{NonlocalInverse}
    \begin{aligned}
        &u=\hat{u}-2\lambda\int dy' G(y+\delta\hat{v}(t,y)-y')\tilde{\MH}_R(y'-t)\\
        &v=\hat{v}-2\lambda\int dy' G(y+\delta\hat{u}(t,y)-y')\tilde{\MH}_L(y'+t)
    \end{aligned}
\end{equation}
which is the same as  \eqref{inverse}, with $v'=y'-t,\,\hat v=y+\delta \hat v-t$.
\end{itemize}
For later convenience, let us define the following field-dependent coordinate, 
\begin{equation}\label{U2s}\begin{aligned}
& U(\sigma^+,\sigma^-) \equiv \sigma^+-2\lambda\int G(\sigma^{-}-v') \tilde{\MH}_R(v')dv'\\   
 &  V(\sigma^+,\sigma^-) \equiv \sigma^--2\lambda\int G(\sigma^+-u') \tilde{\MH}_L(u')du' 
\end{aligned}
\end{equation}
where $\sigma^+, \sigma^-$ are two arbitrary variables. 
Then the inverse transformation \eqref{inverse} or equivalently \eqref{NonlocalInverse} can be written as 
 \begin{align}
&   u=U(\hat u,\hat v),   \quad  v=V(\hat u,\hat v)  \label{U2u}
\end{align}

\subsection{Comment on symmetries on the cylinder}
So far, we have constructed symmetry generators on the plane, and found that they form two copies of the Virasoro algebra \eqref{Witt}. 
Let us now briefly comment on the symmetries of $T\bar T$ deformed CFT$_2$ on the cylinder \cite{Guica:2022gts} as well. 
Despite the similarities, however, there are significant differences between the plane and the cylinder. 
On the cylinder, $T\bar T$ deformation contains an additional $\mathcal Y$ term which deforms the energy spectrum, as shown in \eqref{dHxy}. The finite size effect is also reflected in the symmetry generators. \footnote{Note that our conventions are different from those in \cite{Guica:2022gts}. $u_{here}=U_{there},\hat u_{here}=u_{there}, {\hat u/R_u}|_{here}=\hat u_{there}$. } The fixed period in the physical coordinate $(u,v)\sim (u+ R,v+R)$ leads to a rescaled identification of the non-local coordinates $\hat u$ with $R_u=R+2\lambda H_R$. The Virasoro $\times$ Virasoro generators are then expanded in Fourier modes of the rescaled variable $\hat u/R_u$. 
{The dressed charges on the cylinder, i.e. eq(2.39) in \cite{Guica:2022gts} written in our notation are
\begin{equation}
    \begin{aligned}\label{scaled}
   & \tilde{Q}_n= \frac{R_u}{R} \int^{R/2}_{-R/2}d\sigma e^{\frac{in \hat u}{R_u}} \mathcal{H}_L , \quad \tilde{{\bar Q}}_n= \frac{R_v}{R} \int^{R/2}_{-R/2}d\sigma e^{\frac{in \hat v}{R_v}} \mathcal{H}_R .
\end{aligned}
\end{equation}}
As a result, 
the zero modes of the generators which form the left-moving Virasoro algebra \eqref{Witt} are not the left-moving energy, but rescaled by a factor of $(1+{2\lambda H_R\over R})$. The dressed Virasoro generators are referred to as the rescaled charges in \cite{Guica:2022gts, Guica:2020uhm}, which satisfy the flow equation \eqref{flowedeq} and form two commuting sets of Virasoro algebra. 
On the other hand, the physical Hamiltonian and momentum on the cylinder live in the sets of unrescaled operators, which are obtained from the scaled ones by a factor of $1+{2\lambda H_R\over R}$, 
{\begin{equation}
    \begin{aligned}
   & {Q}_n= \int^{R/2}_{-R/2}d\sigma e^{\frac{in \hat u}{R_u}} \mathcal{H}_L , \quad  {\bar Q}_n=  \int^{R/2}_{-R/2}d\sigma e^{\frac{in \hat v}{R_v}} \mathcal{H}_R .
\end{aligned}\label{unscaled}
\end{equation}}
The unrescaled generators will become a non-linear algebra, with nonvanishing commutators between the left and right-moving sectors.
This has been shown in the Hamiltonian formalism \cite{Guica:2022gts} and Lagrangian formalism \cite{Guica:2020uhm}, and are referred to as the unscaled pseudoconformal symmetry generators. 
Holographically, the rescaled and unrescaled symmetry algebra can be constructed from Einstein gravity on $AdS_3$ with mixed boundary conditions \cite{Guica:2019nzm, Georgescu:2022iyx}, and the string worldsheet \cite{Du:2024bqk} in the TsT/$T\bar T$ correspondence. 
 
In the current paper, we consider the $T\bar{T}$ deformation on the plane, which can be understood as a large $R$ limit of the cylinder case, under which the rescaling factor $1+{2\lambda H_R\over R}\to 1$ so that there is no difference between rescaled and unrescaled symmetries. 
More explicitly, the conformal Killing vectors on the cylinder can be expanded in terms of those on the plane
\begin{equation}
    \ell_n^{\operatorname{cy}} = -Re^{nw\over R}\partial_w= \sum_{m=0}^\infty{R^{1-m} n^{m}\over (m)!} \ell_{m-1}^{\operatorname{pl}}, \quad \ell_{n}^{\operatorname{pl}}=-w^{n+1}\partial_w \label{cl2pl}
\end{equation}
Expand $e^{\frac{in\hat u}{R_u}}$ in the scaled charges \eqref{scaled} and the unscaled charges \eqref{unscaled} on the cylinder, and collect the coefficients of $\hat u^{m+1}$, we obtain their plane limits,
\begin{equation}
\begin{aligned}
    &   \tilde{L}^{\operatorname{pl}}_{m}\equiv\lim_{R\to\infty}\lim_{n\to0}\frac{\p}{\p n^{m+1}}R^{m+1}\tilde{Q}_n=\int^{+\infty}_{-\infty} dy \hat u^{m+1}\mathcal{H}_L\\
&{L}^{\operatorname{pl}}_{m}\equiv\lim_{R\to\infty}\lim_{n\to0}\frac{\p}{\p n^{m+1}}R^{m+1}{Q}_n=\int^{+\infty}_{-\infty} dy \hat u^{m+1}\mathcal{H}_L
\end{aligned}
\end{equation}
We find both the scaled and the unscaled charges on the cylinder in the large radius limit reduce to our dressed symmetry generators \eqref{nonlocalL} on the plane.

\section{The dressed operators} \label{sec: operators}
In this section, we will construct the {dressed} operators explicitly by solving the flow equation \eqref{fO}.

Let us choose the undeformed operators at $t=0$ to be a primary operator with conformal weights $(h,\bar h)$ in the original CFT$_2$, so that we have the Poisson brackets \begin{equation}
\begin{aligned}\label{h0O0}
& \left\{\mathcal{H}_L^{(0)}\left(y_1\right), \mathcal{O}_{h, \bar{h}}^{(0)}\left(y_2\right)\right\}=\left\{H_L^{(0)}, \mathcal{O}_{h, \bar{h}}^{(0)}\left(y_2\right)\right\} \delta\left(y_1-y_2\right)+h \mathcal{O}_{h, \bar{h}}^{(0)}\left(y_2\right) \partial_{y_1} \delta\left(y_1-y_2\right) \\
& \left\{\mathcal{H}_R^{(0)}\left(y_1\right), \mathcal{O}_{h, \bar{h}}^{(0)}\left(y_2\right)\right\}=\left\{H_R^{(0)}, \mathcal{O}_{h, \bar{h}}^{(0)}\left(y_2\right)\right\} \delta\left(y_1-y_2\right)-\bar{h} \mathcal{O}_{h, \bar{h}}^{(0)}\left(y_2\right) \partial_{y_1} \delta\left(y_1-y_2\right)
\end{aligned}
\end{equation}
The above Poisson brackets are preserved by time evolution  \eqref{undeformedO}, and are hence valid at any time.
As discussed in the previous section, the undeformed stress tensor after time evolution does not generate symmetry transformation in any known way. Despite the simplicity of the Ward identity, the physical relevance of the undeformed stress tensor is not clear. It is more informative to examine the transformation of the operators under the deformed stress tensor and the {dressed} stress tensor \eqref{flowed HLR}, the latter of which leads to infinitely many conserved charges \eqref{nonlocalL}.  
After some lengthy calculations, we get the following Ward identities, 
\begin{equation}\label{WI0}
\begin{aligned}
   &i\{\tilde{L}_0,\MOh^{(0)}(y_2)\}=\Big(h+2\lambda\mathcal{B}^{(0)}\Big)\MOh^{(0)}(y_2)+u\p_u \MOh^{(0)} +\p_u\z(\delta\hat{u} \MOh^{(0)}\y)+\p_v\z(\delta\hat{v} \MOh^{(0)}\y)\\
   &i\{\tilde{\bar{L}}_0,\MOh^{(0)}(y_2)\}=\Big(\bar{h}+2\lambda\bar{\mathcal{B}}^{(0)}\Big)\MOh^{(0)}(y_2) +v\p_v\MOh^{(0)}+\p_v\z(\delta\hat{v}\MOh^{(0)}\y)+\p_u\z(\delta\hat{u}\MOh^{(0)}\y)
\end{aligned}
\end{equation}
where 
\begin{equation}
\mathcal{B}^{(0)}=
 \lambda(h+\bar{h}-1) \mathcal{O}_{T \bar{T}}+\frac{2 \bar{h} \mathcal{H}_L}{1+2 \lambda \mathcal{H}}, 
\quad \bar{\mathcal{B}}^{(0)} =  \lambda(h+\bar{h}-1) \mathcal{O}_{T \bar{T}}+\frac{2 {h}  \mathcal{H}_R}{1+2 \lambda\, \mathcal{H}}
\end{equation}
and the derivatives are defined as in \eqref{du}.  When acting on any functional of $\delta \hat u,\delta \hat v$ and local operators $\MOh^{(0)}, \MH_L, \MH_R $, these derivatives are the same as Poisson brackets $\partial_u = \{ \cdot ,H_L \}, \, \partial_v=  \{ \cdot ,H_R \}.$
Except for the terms $\mathcal B^{(0)}$ and $\bar{\mathcal B}^{(0)}$, the right hand side of the Ward identity \eqref{WI0} has been expressed in terms of $\delta\hat u, \delta \hat v, \,\MOh^{(0)}$ and their derivatives. 
In fact, a similar structure also shows up in the flow equations 
\begin{equation}
\begin{aligned}\label{dO0}
\mathcal{D}_\lambda \mathcal{O}_{h, \bar{h}}^{(0)}(y) &={1\over \lambda}\Big(\p_u\big(\delta\hat{u}\mathcal{O}_{h, \bar{h}}^{(0)}\big)+\p_v \big(\delta\hat{v}\mathcal{O}_{h, \bar{h}}^{(0)}\big)\Big)+\Big( {\mathcal{B}}^{(0)}+\bar{\mathcal{B}}^{(0)}\Big)\mathcal{O}_{h, \bar{h}}^{(0)}
\end{aligned}
\end{equation}
where we assume the operator vanishes and $\eta_0,\bar{\eta}_0$ do not diverge as $|y|\to \infty$,
\begin{equation}\label{vanish}
    \mathcal{H}_{L,R}(y),\,\MOh^{(0)}(y)\to 0,\quad \eta_0(y),\bar{\eta}_0(y)\to \text{finite},\quad \text{as} \,\, |y|\to \infty
\end{equation}
The appearance of the same additional terms motivates us to construct a new operator $\MOh$ so that such terms disappear both in the Ward identity and in the flow equations.
 Such an operator will be an important building block to solve the flow equation order by order in $\lambda$.
To this end, we consider dressing the undeformed operator in the following way:
\begin{align}\label{ansatz}
   &\mathcal{O}_{h, \bar{h}} \equiv F^{a} W_{L}^{b} W_{R}^{c} \mathcal{O}_{h, \bar{h}}^{(0)},\nonumber\\
   &F=\left(1+\lambda^2 \mathcal{O}_{T T}\right), \quad W_{L}=\left(1+2 \lambda \mathcal{H}_R\right), \quad W_{R}=\left(1+2 \lambda \mathcal{H}_L\right) .
\end{align}
where the exponents $a,b,c$ are to be determined. The factor $F$ is motivated by the Jacobian in the nonlocal coordinate transformation \eqref{nonlocal0}.
The factors $W_L$ and $W_R$ are motivated by the relation between the deformed and undeformed stress tensor \eqref{RST}, and the nonlocal coordinate transformation \eqref{nonlocal}. The flow equation of these operators and their Poisson brackets with the {dressed} symmetry generators take the same form as in \eqref{WI0} and \eqref{dO0}, 
\begin{equation}\label{Cphyflow}
\begin{aligned}
&\mathcal{D}_\lambda \mathcal{O}_{h, \bar{h}}(y) ={1\over \lambda}\Big(\p_u\big(\delta\hat{u}\mathcal{O}_{h, \bar{h}}\big)+\p_v \big(\delta\hat{v}\mathcal{O}_{h, \bar{h}}\big)\Big)+\Big( {\mathcal{B}}+\bar{\mathcal{B}}\Big)\mathcal{O}_{h, \bar{h}}\\
   &i\{\tilde{L}_0,\MOh\}=\Big(h+2\lambda\mathcal{B}\Big)\MOh+u\p_u \MOh +\p_u\z(\delta\hat{u} \MOh\y)+\p_v\z(\delta\hat{v} \MOh\y)\\
   &i\{\tilde{\bar{L}}_0,\MOh\}=\Big(\bar{h}+2\lambda\bar{\mathcal{B}}\Big)\MOh+v\p_v\MOh+\p_v\z(\delta\hat{v}\MOh\y)+\p_u\z(\delta\hat{u}\MOh\y)
   \end{aligned}
\end{equation}
with the extra factors given by 
\begin{equation}
\begin{aligned}
   &\mathcal{B}=\frac{2(\bar{h}+c)\HL}{1+2\lambda \MH}+\lambda(h+\bar{h}+a+b+c-1) \OT\\
   &\bar{\mathcal{B}}=\frac{2(h+b)\HR}{1+2\lambda \MH}+\lambda(h+\bar{h}+a+b+c-1)\OT
\end{aligned}
\end{equation}
We note that the Lorentzian boost generator acts as
\begin{equation}\label{boostcharge1}
    i\{M, \MOh \}= \z(h-\bar h+ (u\partial_u -v\partial_v ) \y)\MOh +2\lambda (\mathcal{B}-\bar{\mathcal{B}})  \MOh
\end{equation}
In particular, the undeformed operators with $h\neq\bar h$ do not transform nicely under the boost transformation, which implies a better basis of operators are necessary. Both the flow equation and the transformation law simplify greatly when $\mathcal{B}=\bar{\mathcal{B}}=0$, obtained with the choice $ a=1, \, b= -h,  \,  c= -\bar{h}$. In this case, the flow equation becomes a total derivative and the Ward identity with the boost generator remains the same as in a CFT. In fact, these two conditions uniquely fix the above choice within the ansatz \eqref{ansatz}. This motivates us to define an intermediate operator as follows, 
\begin{equation}  
    \mathcal{O}_{h, \bar{h}} \equiv \frac{\left(1+\lambda^2 \, \mathcal{O}_{T \bar{T}}\right)}{\left(1+2 \lambda \mathcal{H}_R\right)^h\left(1+2 \lambda \mathcal{H}_L\right)^{\bar{h}}} \,  \mathcal{O}_{h, \bar{h}}^{(0)}
\end{equation}
We will call this operator the {\it physical} operator, the meaning of which will be further explained later. 
To construct the {dressed} operator, we need to find an operator that vanishes under the action of the covariant derivative $\mathcal D_\lambda$. 
For the physical operator \eqref{physicalOp}, we find that the 
\begin{equation}
\begin{aligned}
& \mathcal{D}_\lambda \mathcal{O}_{h, \bar{h}} ={1\over \lambda}\Big[\partial_u\left(\delta\hat{u} \mathcal{O}_{h, \bar{h}}\right)+\partial_v\left(\delta\hat{v} \mathcal{O}_{h, \bar{h}}\right)\Big] \equiv -{1\over \lambda}{{\tilde {\mathcal O}} }_{h, \bar{h}}^{(1)}
\end{aligned}
\end{equation}
where the right hand side looks very similar to the total differential of a function from $(u,v)$ to   $(u+\delta \hat u,v+\delta \hat v)$, except that the differential element $\delta \hat u, \delta \hat v$ depends on $(u,v)$ and are put inside the derivatives. 
Defining a similar structure at order $n$,
\begin{equation}
  { \tilde {\mathcal O}}_{h, \bar{h}}^{(n)} \equiv (-1)^n \sum_{m=0}^n \frac{1}{(n-m)!m!} \partial_u^m \partial_v^{n-m}\left((\delta\hat{u})^m(\delta\hat{v})^{n-m} \mathcal{O}_{h, \bar{h}}\right)
\end{equation}
we find the following recursion relation,
\begin{equation}
\mathcal{D}_\lambda { \tilde {\mathcal O}}_{h, \bar{h}}^{(n)}=\frac{n}{\lambda} { \tilde {\mathcal O}}_{h, \bar{h}}^{(n)}-\frac{n+1}{\lambda} { \tilde {\mathcal O}}_{h, \bar{h}}^{(n+1)}
\end{equation}
The above recursion relation implies that the following sum is the {dressed} operator we are searching for:
\begin{align}\label{Flowedoperator}
   & \tilde{\mathcal{O}}_{h, \bar{h}} (u,v)\equiv \sum_{n=0}^\infty {\tilde {\mathcal O}}_{h, \bar{h}}^{(n)}= \sum_{n=0}^\infty \sum_{m=0}^{n} \frac{(-1)^n}{(n-m)!m!} \partial_u^m \partial_v^{n-m}\left((\delta\hat{u})^m(\delta\hat{v})^{n-m} \mathcal{O}_{h, \bar{h}}\right)\\
&\mathcal D_\lambda    \tilde{\mathcal{O}}_{h, \bar{h}} =0 \nonumber
\end{align}  Note that the dressed operator \eqref{Flowedoperator} solves the flow equation as long as the operator vanishes and $\eta_0,\,\bar {\eta}_0$ do not diverge as $|y|\to \infty$, as required by \eqref{vanish}.

As reviewed in Sec.\ref{sec: reviews}, the $T\bar{T}$ deformation on the plane is a canonical transformation along the deformation, and the algebra between the {dressed} operators is preserved along the deformation. Indeed, we can explicitly check that
the {dressed} operators \eqref{Flowedoperator} are primary operators with conformal weights $(h,\bar{h})$ under the symmetry generators \begin{equation}
\begin{aligned}
   &i\{\tilde{L}_m,\tilde{\MO}_{h,\bar{h}}\}=(m+1)hx^m\tilde{\MO}_{h,\bar{h}}+x^{m+1}\p_{x}\tilde{\MO}_{h,\bar{h}}\\
   &i\{\tilde{\bar{L}}_m,\tilde{\MO}_{h,\bar{h}}\}=(m+1)\bar{h}\bar{x}^m\tilde{\MO}_{h,\bar{h}}+\bar{x}^{m+1}\p_{\bar{x}}\tilde{\MO}_{h,\bar{h}}
\end{aligned}
\end{equation}
 where $x=iu$ and $\bar{x}=-iv$ are lightcone coordinates in Euclidean space.

 Consider an arbitrary function $f(u,v)$ that vanishes fast enough at infinity. 
By using integration by parts, we notice that the following relation from \eqref{Flowedoperator}, 
\begin{equation}\label{inteO}
\int dudvf(u,v)\tilde{\MO}_{h,\bar{h}}(u,v)=\int dudv f(\hat{u},\hat{v})    \MO_{h,\bar h}(u,v)
\end{equation}
where the nonlocal coordinates are defined in \eqref{nonlocal}. Similar to the case of dressed stress tensor, by choosing different $f$, the above relation can be used to derive useful relations between the dressed and physical operators.
\begin{itemize}
    \item 
By choosing $f(u,v)=e^{-ip_Lu+ip_Rv}$, we have the momentum space operator
\begin{equation}
\begin{aligned}
   \tilde{\MO}_{h,\bar{h}}(p_L,p_R)&\equiv \mathcal{F}_{p_L,p_R}[\mathcal{\tilde O}_{h, \bar{h}}(u,v)]  =\int du dv e^{-ip_Lu+ip_Rv}\tilde{\MO}_{h,\bar{h}}(u,v)\\
   &=\int dudv e^{-ip_L\hat{u}+ip_R\hat{v}}\MOh (u,v)
\end{aligned}
\end{equation}
where $\mathcal{F}_{p_L,p_R}[\cdot]$ denotes taking Fourier transformation. In other words, the dressed operator in momentum space which is by definition the Fourier transformation of the dressed operator with respect to coordinates $(u,v)$, can be equivalently understood as decomposing the physical operator into plane wave basis in the nonlocal coordinates $(\hat u,\hat v)$. 
\item Consider the function 
\begin{equation}
f(u',v')=\delta(u'-\hat{u}(u),v'-\hat{v}(v)) 
\end{equation}
and we obtain the formal relation between the physical and the {dressed} operator
\begin{equation}\label{relation}
\tilde{\MO}_{h,\bar{h}}(\hat{u},\hat{v})={\MOh(u,v)\over (1+\lambda^2\OT)}\end{equation}
   can be viewed as a function of $(u,v)$. The left-hand side of the above formula is obtained from the dressed operator \eqref{Flowedoperator}-viewed as a function of $(u,v)$- by replacing the variables $(u,v)$ by $\hat u,\,\hat v$. 
This provides a simple relation without the infinite sum: the dressed operator living on an auxiliary space with coordinates $(\hat u, \hat v)$ is related to a local expression of the physical operator living on $(u,v)$. 

\item Furthermore, using the relation between the physical and undeformed operator, we find 
\begin{equation}\label{O0relation}
\tilde{\MO}_{h,\bar{h}}(\hat{u},\hat{v})={\left(1+2 \lambda \mathcal{H}_R\right)^{-h}\left(1+2 \lambda \mathcal{H}_L\right)^{-\bar{h}}} \,  \mathcal{O}_{h, \bar{h}}^{(0)}(u,v)
\end{equation}
This provides another expression for the {dressed} operator without using the infinite sum, 
\begin{equation}
\tilde{\MO}_{h,\bar{h}}({u},{v})={\Big(1-4\lambda^2\tilde{\mathcal{H}}_L(u)\tilde{\mathcal{H}}_R(v) \Big)^{h+\bar h}\over \left(1+2\lambda \tilde{\mathcal{H}}_R(v)\right)^{h}\left(1+2 \lambda \tilde{\mathcal{H}}_L(u)\right)^{\bar{h}}} \,  \mathcal{O}_{h, \bar{h}}^{(0)}(U(u,v),V(u,v))
\end{equation}
where we have used the inverse non-local coordinate transformation \eqref{U2s}, and the relation \eqref{Hrelation}. The above relation expresses the dressed operator living on the physical space $(u,v)$ to undeformed operators that depend on nonlocal variables $U,\,V$.
\end{itemize}

To recapitulate, we have constructed dressed operators explicitly by solving the flow equation. The result can be written in the form of an infinite sum \eqref{Flowedoperator}, or in terms of operators on nonlocal variables \eqref{relation} and \eqref{O0relation}.

\subsubsection*{Comments on the stress tensor}

So far, the derivation of the {dressed} operator in this section is purely classical, and it should be valid for any primary operators in the undeformed theory which satisfy the Poisson bracket \eqref{h0O0}.  Assuming no classical central charge at the classical level, the undeformed stress tensor also satisfies the Poisson bracket \eqref{h0O0}.  Then we need to understand if the {dressed} operator \eqref{Flowedoperator} constructed in this section is consistent with \eqref{flowed HLR} in the previous section, and furthermore, if the deformed stress is related to the so-called physical operator in this section. In order to address these questions, we can take \begin{align}\mathcal O^{(0)}_{2,0}=\mathcal H_L^{(0)}, \quad \mathcal O^{(0)}_{0,2}=\mathcal H_R^{(0)}.\end{align} Then it is clear that the relation between the {dressed} and undeformed stress tensor \eqref{H0relation} takes the same form as \eqref{O0relation}, and thus the two ways of constructing the {dressed} stress tensor are indeed compatible.

However, comparing the relation \eqref{RST} and  \eqref{physicalOp}, the deformed stress tensor  is not the physical operator, but is related by
\begin{align}\label{o20}
\mathcal O_{2,0}={1+\lambda^2\OT \over  1+2 \lambda \mathcal{H}_R}\mathcal H_L\end{align}
Plugging this into \eqref{Flowedoperator},
\begin{align}\label{secondHL}
   & \tilde{\mathcal{O}}_{2, 0} (u,v)= \sum_{n=0}^\infty \sum_{m=0}^{n} \frac{(-1)^n}{(n-m)!m!} \partial_u^m \partial_v^{n-m}\left((\delta\hat{u})^m(\delta\hat{v})^{n-m} \z({1+\lambda^2\OT \over  1+2 \lambda \mathcal{H}_R}\mathcal H_L\y)\right)
\end{align}
The above expression looks different from \eqref{flowed HLR}, and it is difficult to directly verify the equivalence.
Nevertheless, plugging \eqref{o20} into \eqref{relation}, we find the expression of the {dressed} $(2,0)$ operator in terms of the nonlocal coordinate is given by
\begin{equation}
 \tilde{\mathcal{O}}_{2, 0} (u,v)={\MH_L(u,v)\over 1+2\lambda \MH_R(u,v) }= \tilde{\MH}_L(\hat{u})
\end{equation}
which is consistent with \eqref{Hrelation} derived from \eqref{flowed HLR}. This implies \eqref{secondHL} and \eqref{Hrelation} indeed define the same {dressed} stress tensor.

In this subsection, we have constructed the {dressed} primary operators \eqref{Flowedoperator} in the classical $T\bar T$ deformed CFT.  We have also shown equivalent expressions \eqref{inteO} in terms of integrals, and  \eqref{O0relation} in terms of nonlocal coordinates.
 With an appropriate normal ordering prescription, we expect that these operators exist at the quantum level, and their correlation functions remain the same as in the undeformed CFT$_2$.

\section{The physical operators}

In the derivation of the {dressed} operators, we have encountered an intermediate operator  \eqref{physicalOp}, which we reproduce here for convenience, 
\begin{equation}  \label{physicalOp2}
    \mathcal{O}_{h, \bar{h}} \equiv \frac{\left(1+\lambda^2 \, \mathcal{O}_{T \bar{T}}\right)}{\left(1+2 \lambda \mathcal{H}_R\right)^h\left(1+2 \lambda \mathcal{H}_L\right)^{\bar{h}}} \,  \mathcal{O}_{h, \bar{h}}^{(0)}
\end{equation}
This operator is by definition a local expression in terms of fundamental fields, and referred to as the {\it physical } operator. 
In section 4, we have expressed the dressed operator in terms of physical operators. In this section, however, we would like to invert the relation and find an expression of physical operators in terms of the dressed ones, so that conformal symmetry can still be used to organize calculations involving the physical operators. 

In the following, we will first examine the relation between physical operators and dressed operators, and then compare the physical operator defined here to the operator discussed in \cite{Aharony:2023dod}. We end this section with the example of free boson theory. 

\subsection{Physical operators from the dressed operators}
We have derived a useful relation \eqref{inteO} between the physical operators and dressed operators, which we reproduce here for convenience, 
\begin{equation}\label{Flow2phys}
\int dudvf(u,v)\tilde{\MO}_{h,\bar{h}}(u,v)=\int dudv f(\hat{u},\hat{v})    \MO_{h,\bar h}(u,v)
\end{equation}
Taking the function $f(u,v)=\delta(U(u,v)-u_0,V(u,v)-v_0)$, where $U, V$ are the nonlocal functions defined in \eqref{U2s},  it follows that $ f(\hat u,\hat v)=\delta(u-u_0,v-v_0) $. Plugging into the  
 \eqref{Flow2phys} we have 
\begin{align}
\MOh(u_0,v_0)=\int dudv\delta(U(u,v)-u_0,V(u,v)-v_0)\tilde{\MO}_{h,\bar{h}}(u,v)\label{Ophys1}
\end{align}
We can rename $u,v$ on the right hand side as $\sigma^\pm$ are dummy variables, and 
drop the subscripts of $(u_0,v_0)$ as they are arbitrarily chosen, so that we obtain a formal expression of the physical operator in terms of the {dressed} ones, 
\begin{align}
\MOh(u,v)=\int d\sigma^2\delta(U(\sigma^+,\sigma^-)-u,V(\sigma^+,\sigma^-)-v)\tilde{\MO}_{h,\bar{h}}(\sigma^+,\sigma^-)\label{Ophys}
\end{align}
Note that the delta function on the right-hand side localizes the integral to the points at which the equation $ U(\sigma^+,\sigma^-)=u$ is satisfied. From  \eqref{U2u} we learn that the solution is  \footnote{ We assume that there are no other solutions to $ U(\sigma^+,\sigma^-)=u$. } \begin{equation}\sigma^+=\hat u(u,v), \quad \sigma^-=\hat v(u,v), \label{onshell}\end{equation} so that \eqref{Ophys} can be integrated out and we obtain 
\begin{align}
\MOh(u,v)=\det({\partial X\over \partial \sigma})^{-1} \tilde{\MO}_{h,\bar{h}}(\hat u,\hat v)={\tilde{\MO}_{h,\bar{h}}(\hat u,\hat v)\over 1-4\lambda^2 \tilde \MH_L(\hat u)\tilde \MH_R(\hat v)} \label{Ophys2}
\end{align}
where $\det({\partial X\over \partial \sigma})$ denotes the Jacobian in the change of variables \eqref{U2s}.
As a consistent check,  we find the above relation between the physical operator and {dressed} operator is precisely the same as \eqref{relation}.  This can be verified by using the relation between the {dressed} stress tensor and deformed stress tensor \eqref{Hrelation}  and the expression \eqref{Ottbar}. 

We can also perform a Fourier transform by 
plugging \begin{align}
&f(u,v) =e^{-ip_L U(u,v)+ip_R V(u,v)},   \quad f(\hat{u}, \hat{v}) = e^{-i p_L u+ i p_R v}
\end{align}
 into \eqref{Flow2phys},  so that the physical operator  in  momentum space is given by 
 \begin{equation}\label{TTbardress}
\begin{aligned}
\MOh(p_L,p_R)
&=\int dudv  e^{ -ip_L U(u,v)+ip_R V(u,v)} \tilde{\MO}_{h,\bar{h}}(u,v) 
\\
&=\mathcal{F}_{p_L,p_R}[e^{ip_L\z( 2\lambda \int G(v-v') \tilde{\MH}_R(v')dv'\y)-ip_R\z(2\lambda\int G(u-u')\tilde{\MH}_L(u')du'\y)}\tilde{O}_{h,\bar{h}}(u,v)]
\end{aligned}
\end{equation}

We can write down the full Ward identities of these operators under the {dressed} symmetry generators,
\begin{equation}
\begin{aligned}
   i\{\tilde{L}_m,\MOh\}&=i^m(m+1)\Big(h+2\lambda^2 \OT\Big)(u+\delta\hat{u})^m\MOh+i^m(u+\delta\hat{u})^{m+1}\p_u\MOh\\
   &+2\lambda (m+1)\eta_m\p_v\MOh\\
   i\{\tilde{\bar{{L}}}_m,\MOh\}&=(-i)^m(m+1)\Big(\bar{h}+2\lambda^2 \OT\Big)(v+\delta\hat{v})^m\MOh+(-i)^m(v+\delta\hat{v})^{m+1}\p_v\MOh\\
   &+2\lambda (m+1)\bar{\eta}_m\p_u\MOh
\end{aligned}
\end{equation}
where $\eta_m$ is defined in \eqref{etam}. 
The Ward identity in momentum space is given by 
\begin{equation}\label{TTbarWard}
\begin{aligned}
   i\{\tilde{L}_m,\MOh(p_L,p_R)\}&= \sum_{k=0}^m \frac{(-1)^k (m+1)!}{k!(m-k)!}\left[(h-1) -\frac{p_L}{k+1} \frac{\partial}{\partial p_L}\right] \frac{\partial^k}{\partial p_L^k} \mathcal{F}_{p_L, p_R}\left[(i \delta \hat{u})^{m-k} \mathcal{O}_{h,\bar{h}}\right]\\
   &+p_L \mathcal{F}_{p_L, p_R}[(i \delta \hat{u})^{m+1} \mathcal{O}_{h,\bar{h}}] -2\lambda(m+1)p_R\mathcal{F}_{p_L,p_R}[i\eta_m\MOh]\\ \\
   i\{\tilde{\bar{L}}_m,\MOh(p_L,p_R)\} &=  \sum_{k=0}^m \frac{(-1)^k (m+1)!}{k!(m-k)!}\left[(\bar{h}-1) -\frac{p_R}{k+1} \frac{\partial}{\partial p_R}\right] \frac{\partial^k}{\partial p_R^k} \mathcal{F}_{p_L, p_R}\left[(-i \delta \hat{v})^{m-k} \mathcal{O}_{h, \bar{h}}\right] \\ 
& +p_R \mathcal{F}_{p_L, p_R}\left[(-i \delta \hat{v})^{m+1} \mathcal{O}_{h, \bar{h}}\right]+2 \lambda(m+1) p_L \mathcal{F}_{p_L, p_R}\left[i \bar{\eta}_m \mathcal{O}_{h, \bar{h}}\right]
\end{aligned}
\end{equation}
where $m \geq -1$. From the above Ward identity, we note that the momentum space operator is an eigenstate of $\tilde{L}_{-1}$ with eigenvalue $p_L$.
It also has fixed spin, as can be seen from the Ward identity with the boost generator in the position space \eqref{boostcharge1}, or in momentum space \begin{align} i\{ M, \MOh(p_L,p_R)\}\}=\z((h-\bar h)-p_L \partial_{p_L}+p_R \partial_{p_R}\y)\MOh(p_L,p_R)\label{boostchargeM}\end{align}

\subsection{Comparison to alternative definitions of operators }
In this subsection, we compare the physical operator \eqref{Ophys} with the operator in the computation of correlation functions \cite{Aharony:2023dod}. We find that at the classical level, these operators are the same. 
 
Let us first review the definition of operator in \cite{Aharony:2023dod}. The $T\bar{T}$ deformed theory is reformulated in terms of topological gravity \cite{Dubovsky:2017cnj,Dubovsky:2018bmo}, 
\begin{equation}\label{top}
\begin{aligned}
S_{T\bar{T}}\left(\psi, e_\alpha^a, X^a\right) & =S_0\left(\psi, e_\alpha^a\right)+S_{grav}\left(e_\alpha^a, X^a\right) \\
& =S_0\left(\psi, e_\alpha^a\right)+\frac{\Lambda}{2} \int d^2 \sigma \epsilon^{\alpha \beta} \epsilon_{a b}\left(\partial_\alpha X^a-e_\alpha^a\right)\left(\partial_\beta X^b-e_\beta^b\right)
\end{aligned}
\end{equation}
 where $S_0(\psi,\delta_\alpha^a)$ is the Lagrangian of the undeformed CFT on the plane, with vielbein $e^{a}_{\alpha} $ and minimal coupled matter field $\psi$,   $X^{a}  $ are the coordinates of the target space. After integrating out $e^a_\mu$ and $X^a$ one  gets the $T\bar{T}$ deformed  theory with deformation parameter $\lambda=\Lambda 
^{-1}$.
 In this context, the authors in \cite{Aharony:2023dod} considered local operators in $T\bar{T}$ deformed theory as follows:
    \begin{equation} \label{AharonyO}
        \mathcal{O}^{AB}(u,v) \equiv \int d^2 \, \sigma  \sqrt{g(\sigma^a)} \, \mathcal{O}^{C F T}(\sigma) \, \delta^2(U(\sigma^a)-u, V(\sigma^a)-v)\\
          \end{equation}
where $T_{\pm\pm}\equiv T_{\sigma_\pm\sigma_\pm}$ are left and right-moving stress tensor of the worldsheet CFT  $S_0$.    
In this formulation, the operators are defined on the target spacetime with coordinates $x^a=(u,v)$, where the $T\bar{T}$ deformed theory lives on. $U(\sigma)$ are scalar fields on worldsheet with coordinates $\sigma^\alpha$ and metric $g_{\alpha\beta}=e_\alpha^ae_{a\beta}$. The delta function in the definition \eqref{AharonyO} then relates the dynamical coordinates $X^a(\sigma)$ to the target spacetime coordinates $x^a$. 

At the classical level, we can solve the equations of motion for $e^a_\alpha$ and $X^a$
\begin{equation}
	\begin{aligned}
		&\epsilon^{\alpha\beta}\p_\alpha e_\beta^a=0\\
		&\p_\beta X^b=e^b_\beta+\lambda\epsilon^{ab}\epsilon_{\alpha\beta}\det(e^a_\alpha)T_a^\alpha	\end{aligned}\label{Xeq}
\end{equation}
where $T_a^\alpha\equiv \det(e^a_\alpha)^{-1}\frac{\delta S_0(\psi,e^{a}_\alpha)}{\delta e^a_\alpha}$ is the worldsheet stress tensor. 
Using lightcone coordinates on the worldsheet, the first equation can be solved by 
\begin{equation}
	\begin{aligned}
		&e_{\sigma_\pm}^{a}=e_{\sigma_\pm}^{a}(\sigma_\pm)
	\end{aligned}
\end{equation}
Using the diffeomorphism and local Lorentzian gauge symmetry on the worldsheet, we can make a gauge choice of $e^a_{\alpha}=\delta^a_\alpha$. Then  the dynamical lightcone coordinates $X^a=(U,V)$  satisfy 
\begin{equation}\label{UV2sigma}
	\begin{aligned}
	&\p_{\sigma^+}U=1,\quad \p_{\sigma^-}U=-2\lambda T_{--}(\sigma^-)\\
	&\p_{\sigma^+}V=-2\lambda T_{++}(\sigma^+),\quad \p_{\sigma^-}V=1
	\end{aligned}
\end{equation}
The solution to the above equations is
\begin{equation}\begin{aligned}\label{U2ss}
& U(\sigma^+,\sigma^-) \equiv \sigma^+-2\lambda\int G(\sigma^--v') T_{--}(v')dv',   \\
 &  V(\sigma^+,\sigma^-) \equiv \sigma^--2\lambda\int G(\sigma^+-u') T_{++}(u')du' 
\end{aligned}
\end{equation}
where $u',v'$ are null coordinates on the plane. 
The above solution is just \eqref{U2s}, with $\sigma^\pm$ now understood as the worldsheet coordinate, and $U,V$ as the dynamical coordinates, and $\tilde H_L$ identified as the $T_{++}$.  
Then the delta function in the definition of \eqref{AharonyO} localizes at the worldsheet points which solve the equation $U(\sigma^a)=u$, which are non-local in terms of coordinates $x^a=(u,v)$ given by  \eqref{onshell}.
In other words, we have the following on-shell relations
\begin{align}\label{onshelluv}
&\sigma^+ = \hat u(u,v), \quad \sigma^- = \hat v(u,v)\Longleftrightarrow  U(\sigma^+,\sigma^-)= u,\quad V(\sigma^+,\sigma^-)= v
\\
&T_{++}(\sigma^+)= \tilde \MH_L(\hat u), \quad T_{--}(\sigma^+)= \tilde \MH_R(\hat v),\nonumber \end{align}
This also suggests that we identify the CFT operators with the {dressed} operators, as
\begin{align} O^{CFT}(\sigma^+,\sigma^-)\leftrightarrow \tilde O(\hat u,\hat v) ={{\mathcal O}^{(0)}(u,v)\over (1+2\lambda \mathcal H_R)^h(1+2\lambda \mathcal H_L)^{\bar h}}\label{OcftOt}
\end{align}
where we have used \eqref{O0relation} in the last expression. 
Then the physical operator \eqref{Ophys} we defined indeed agrees with \eqref{AharonyO} at the classical level, \begin{align}
\mathcal O(u,v)=\mathcal O^{AB}(u,v)
\end{align}

 \subsection{Example: Operators in $T\bar{T}$ deformed free scalar theory}
  In this subsection, we illustrate the equivalence between the physical operators \eqref{Ophys} and  \eqref{AharonyO} in a simple example.
We start with the free scalar theory before the $T\bar T$ deformation with the Hamiltonian density and momentum density
\begin{equation}
	\begin{aligned}
		\MH^{(0)}=\frac{1}{2}(\pi_\phi^2+\p_y\phi\p_y\phi)
	\quad 
	\MP=\pi_\phi\p_y\phi\end{aligned},
\end{equation}
with the canonical commutation relation \begin{equation}   \{\phi(y), \pi_\phi(y')\}=\delta(y-y') \end{equation}
According to \eqref{deformedH}, the deformed Hamiltonian density is
\begin{equation} \label{scalarSpectrum}
	\mathcal{H}=\frac{1}{2\lambda}\z(\sqrt{(1+2\lambda \, \pi_\phi^2)(1+2\lambda \,(\p_y\phi)^2)}-1\y)
\end{equation}
\subsubsection*{The physical operator}
Let us consider a primary operator with dimension $(1,0)$ in the original CFT,
\begin{equation}
	\MO^{CFT}_{1,0}=\p_u\phi=\frac{1}{2}(\pi_\phi+\p_y\phi)\label{Ocft}
\end{equation}
where we have used the equation of motion $\p_t\phi\equiv \{\phi,H^{(0)} \} = \pi_{\phi} $ in the undeformed CFT to express the operator in Hamiltonian formalism. 
This allows us to define the undeformed operator \eqref{undeformedO} in the $T\bar T$ deformed CFT as:
\begin{equation}
	\MO^{(0)}_{1,0}=\frac{1}{2}(\pi_\phi+\p_y\phi)\label{O0scalar}
\end{equation}
Although the undeformed operator \eqref{O0scalar} in the $T\bar T$ deformed theory has the same expression as the CFT operator \eqref{Ocft} in terms of the canonical variables, they differ in the Lagrangian formalism as the relation between time derivative and canonical momentum is different. 
Using the equation of motion in the deformed theory, we get 
\begin{equation}
	\p_t\phi=\{\phi,H\}=\frac{\pi_\phi(1+2\lambda \, (\p_y\phi)^2)}{\sqrt{(1+2\lambda \, \pi_\phi^2)(1+2\lambda \,(\p_y\phi)^2)}}
\end{equation}
so that the canonical momentum  $\pi_{\phi}$ is given by
 $\p_t\phi$ 
\begin{equation}
	\pi_\phi=\frac{\p_t\phi}{\sqrt{1+8\lambda\p_u\phi\p_v\phi }}
\end{equation}
As a consequence, the undeformed operator written in the Lagrangian formalism is 
\begin{equation}
	\MO^{(0)}_{1,0}=\frac{1}{2}\z(\frac{\p_t\phi}{\sqrt{1+8\lambda\p_u\phi\p_v\phi }}+\p_y\phi\y)\label{O0scalarL}
\end{equation}
which actually depends on $\lambda$.
Using the relation between the physical operator and undeformed operator \eqref{physicalOp}, the physical operator  becomes 
\begin{equation}\label{Ophyss}
	\MO_{1,0} (u,v)\equiv\frac{(1+\lambda^2\OT)}{(1+2\lambda \HR)} {\mathcal O}_{1,0}^{(0)}=\frac{\p_u\phi\z(1+\sqrt{1+8\lambda\p_u\phi\p_v\phi}\y)}{2\sqrt{1+8\lambda\p_u\phi\p_v\phi}}\,.
\end{equation}
The {dressed} operator can be written as 
\begin{equation} {\tilde O}_{1,0}(\hat u, \hat v)={{\mathcal O}^{(0)}_{1,0}\over (1+2\lambda \HR)}= \frac{2\p_u \phi}{1+\sqrt{1+8\lambda \p_u\phi\p_v\phi}}  \end{equation}

\subsubsection*{The Aharony-Barel operator }
Now let us consider the operator \eqref{AharonyO} in the $T\bar T$ deformed theory \eqref{top} with the undeformed theory $S_0(\phi,\delta^a_\alpha)$ being the free scalar theory.
The operator \eqref{AharonyO} is a local operator on the target spacetime coordinate $(u,v)$ but expressed as a worldsheet integral. 
To perform the integral, we need to perform the nonlocal coordinate transformation \eqref{U2ss}, and then identify the dynamical coordinate $X^a(\sigma)$ as the spacetime coordinate  $x^a.$ 
Using the nonlocal coordinate transformation \eqref{UV2sigma} and the worldsheet stress tensor 
  \begin{equation} \label{freeStressT}
    T_{++}=\p_{\sigma^+}\phi\p_{\sigma^+}\phi,\quad T_{--}=\p_{\sigma^-}\phi\p_{\sigma^-}\phi
\end{equation}
 we  obtain \begin{equation}
    \begin{aligned}
        &\p_{\sigma^+}\phi=\p_U\phi-2\lambda\p_{\sigma^+}\phi\p_{\sigma^+}\phi\p_V\phi\\
        &\p_{\sigma^-}\phi=\p_V\phi-2\lambda\p_{\sigma^-}\phi\p_{\sigma^-}\phi\p_U\phi
    \end{aligned}
\end{equation}
The above equations can be viewed as quadratic equations for $\p_{\sigma^+}\phi$, so that the CFT operators can be written in the nonlocal coordinate as 
\begin{equation}\label{dsphi}
    O_{1,0}^{CFT}(\sigma^a)\equiv\p_{\sigma^+}\phi = \frac{ 2 \p_U \phi}{1+\sqrt{1+8\lambda \p_U\phi\p_V\phi}},\quad O_{0,1}^{CFT}(\sigma^a)\equiv  \p_{\sigma^-}\phi= \frac{ 2 \p_V \phi}{1+\sqrt{1+8\lambda \p_U\phi\p_V\phi}}
\end{equation}
Here we have chosen the branch of solution that returns to the undeformed theory in the limit $\lambda \to0$. Performing the integral of \eqref{AharonyO}, we obtain the operator 
    \begin{equation}
        \begin{aligned}    \mathcal{O}^{AB}_{1,0}(u,v)&=\z[\operatorname{det}^{-1}\left(\frac{\partial X}{\partial \sigma}\right) \partial_{\sigma^+} \phi(\sigma) \y] {\bigg|}_{U,V=u,v} =\frac{\p_u\phi\z(1+\sqrt{1+8\lambda\p_u\phi\p_v\phi}\y)}{2\sqrt{1+8\lambda\p_u\phi\p_v\phi}}={\mathcal{O}}_{1,0}(u,v)
        \end{aligned}
    \end{equation}
which is indeed the same as the physical operator $\MO_{1,0}$ \eqref{Ophyss}. 
Using the onshell condition \eqref{onshelluv}, we also find 
\begin{align}
O_{1,0}^{CFT}(\hat u,\hat v)=O_{1,0}^{CFT}(\sigma^+,\sigma^-)|_{U=u,V=v}= \frac{ 2 \p_u \phi}{1+\sqrt{1+8\lambda \p_u\phi\p_v\phi}}={\tilde O}_{1,0}(\hat u, \hat v)
\end{align}
which further justifies the identification of the {dressed} operator with the CFT operator upon the on-shell condition.

In the example of free scalar, we have shown explicitly that the physical operator $\MO_{1,0}$ \eqref{Ophyss} is equivalent to the operator defined in \cite{Aharony:2023dod} at the classical level. In the next section, we will further show that after uplifting our physical operator to the quantum level,  their correlation functions can also be matched to the calculation of \cite{Aharony:2023dod}.

\section{Correlation functions}
In the previous section, we have defined the physical operator in the $T\bar T$ deformed CFTs at the classical level, which is local but expressible in terms of the {dressed} operators. The latter properties enable us to leverage conformal symmetry to compute physical quantities.  In this section, we will first promote the physical operator to the quantum level, and then compute the correlation functions in the $T\bar T$ deformed CFTs. 

\subsection{Physical operator at the quantum level}

At the classical level, the physical operators can be expressed in two equivalent ways: as a local expression in terms of the undeformed operator \eqref{physicalOp} and as an integral expression in terms of the dressed operator \eqref{Ophys2} in position space and \eqref{TTbardress} in momentum space. 
At the quantum level, it is not clear if all the expressions are equivalent, and we need to define the physical operator by promoting one of the classical expressions with proper normal ordering prescription. While the relation with the undeformed operator \eqref{physicalOp} is useful in arguing its local property at the classical level, the appearance of the operators in the denominator makes it difficult to promote to the quantum level. Instead, the expression \eqref{TTbardress} is a better starting point. To proceed, we propose the following prescription.

We assume the existence of the dressed operators at the quantum level, which satisfy the flow equations and has a classical expression in terms of the undeformed operator as \eqref{Flowedoperator}. In particular, the
Ward identities between the dressed stress tensor and dressed primary operators, and correlation functions between dressed operators remain the same as in a CFT.
Using the dressed operators as the building blocks, we promote the expression \eqref{TTbardress} as the definition of the physical operator. 
 More explicitly, we define the physical operator in position space 
\begin{align}
\MOh(u,v)=\int d\sigma^2\delta(U(\sigma^+,\sigma^-)-u,V(\sigma^+,\sigma^-)-v)\tilde{\MO}_{h,\bar{h}}(\sigma^+,\sigma^-)\label{OphysQ}
\end{align}
which is the same as \eqref{Ophys} but assuming normal ordering on the right-hand side. 
In the momentum space, the physical operator is defined as
\begin{equation}
\begin{aligned}\label{OphypQ}
\MOh(p_L,p_R)
&\equiv \int du dv \, e^{ ip_{L} u - ip_{R} v} \, \Big[e^{i2\lambda \Big(p_L \int^{v} \tilde{\MH}_R(v')dv'-p_R \int^{u} \tilde{\MH}_L(u')du' \Big)} \, \tilde{O}_{h,\bar{h}}(u,v) \Big]
\end{aligned}
\end{equation}
which is the same as \eqref{TTbardress} but assuming normal ordering on the right-hand side.

As discussed in \cite{Cavaglia:2016oda,Brizio:2024arr}, the precise definition of a composite operator depends crucially on the explicit normal ordering prescription. 
As we will show momentarily, we will use point-splitting in Euclidean space to further sharpen the definition. We will further comment on the definition in the next section. 


\subsection*{The flow equation in position space}
 Before moving on with explicit computation of the correlation functions, let us consider the flow equations which we discussed for the undeformed operators in \cite{Cardy:2019qao,Kruthoff:2020hsi}.
Using the relation \eqref{Ophys} between physical operators and dressed operators, we have 
\begin{equation}
\begin{aligned}
        \p_\lambda\left\langle \MO_{h,\bar{h}}(u,v)... \right\rangle=&\int d\sigma^2\Big\langle \Big(2 \int^{\sigma^-}dv'\tilde{\MH}_R(v')\p_u\delta(U(\sigma^+,\sigma^-)-u,V(\sigma^+,\sigma^-)-v)\tilde{\MO}_{h,\bar{h}}(\sigma^+,\sigma^-)\\
    &+2 \int^{\sigma^+}du'\tilde{\MH}_L(u')\p_v\delta(U(\sigma^+,\sigma^-)-u,V(\sigma^+,\sigma^-)-v)\tilde{\MO}_{h,\bar{h}}(\sigma^+,\sigma^-)\Big)... \Big\rangle
\end{aligned}
\end{equation}
where we have used the fact that the correlators between dressed operators do not flow, and hence the derivative only acts on the explicit dependence of $\lambda$.
Using the relation between dressed stress tensor and the deformed one \eqref{Ginverse}, the flow equation can be written as
\begin{align}\label{floweq}
    \partial_\lambda\left\langle \mathcal{O}_{h,\bar{h}}(u,v)\dots \right\rangle 
    &= \left\langle \epsilon^{ab}\epsilon_{ij}\partial_b \left( \int^{(u,v)} \! dx^{\prime j} \, T_{a}{}^i(x') \mathcal{O}_{h,\bar{h}}(u,v) \right) \dots \right\rangle \\
    &= \left\langle \epsilon^{ab}\epsilon_{ij} \left( \int^{(u,v)} \! dx^{\prime j} \, T_{a}{}^i(x') \partial_b \mathcal{O}_{h,\bar{h}}(u,v) \right) \dots \right\rangle 
    + \left\langle  T_a\,^a \, \mathcal{O}_{h,\bar{h}}(u,v) \dots \right\rangle \nonumber
\end{align}
At first glance, the above flow equation appears to be similar to that of \cite{Cardy:2019qao} and \cite{Kruthoff:2020hsi}\footnote{See eq.(3.14) of \cite{Cardy:2019qao}, and eq.(3.7) of \cite{Kruthoff:2020hsi}. }. The difference, however, is that the derivative in \eqref{floweq} is a total derivative, whereas in \cite{Cardy:2019qao, Kruthoff:2020hsi}, the derivative only acts on the operator itself, i.e. the first term in the second line. It is interesting to note that the difference is a universal term that depends on the trace of the stress tensor.
This is not surprising, as we are considering a different type of operator from those considered in the aforementioned two papers. The distinction is most obvious when comparing   \cite{Kruthoff:2020hsi} and our paper, as both are formulated in the Hamiltonian formalism. \cite{Kruthoff:2020hsi} considers the flow of undeformed operators, and we consider the physical operators. 
Many perturbative results \cite{Kraus:2018xrn,He:2020udl,He:2023kgq} of correlation functions in $T\bar T$ deformed CFTs are also similar to those of the undeformed operators. 

\subsection*{Series expansion}  
For convenience, we will compute the correlation function in Euclidean signature, with the following Wick rotation,
\begin{align}u\to -ix, \,v \to i\bar x,\quad p_L\to  -ip, \, p_R\to -i\bar p,  \quad \tilde{\MH}_L(u) \to {T(x)\over 2\pi}, \quad \tilde{\MH}_R(v) \to {\bar T(\bar{x}) \over 2\pi}\end{align}
Then the physical operator in momentum space becomes:
\begin{equation}
\begin{aligned}\label{OE}
		\MOh(p,\bar{p})&\equiv\int dx^2 \,e^{ipx+i\bar{p} \bar{x}} \Big[ e^{i{\lambda\over\pi}\Big( p{ \int^{\bar{x}} \bar{T}(\bar{x}')d\bar{x}'}  +\bar{p}  \int^{{x}} {T}({x}')d{x}'  \Big)} \, \tilde{\MO}_{h,\bar{h}}(x,\bar{x})\Big]
	\end{aligned}
\end{equation}
The operator in the square bracket in the above expression looks like a Wilson line dressing of the {dressed} operator. 
As discussed in section 2, the {dressed} operator $\tilde{\MO}_{h,\bar{h}}(x,\bar{x})$ satisfies the Ward identities under the {dressed} stress tensor $T(x),\,\bar T(\bar x)$, and has the same correlation function as in the undeformed CFT$_2$. This enables us to compute the correlation functions between the physical operators by expanding the  Wilson-line like dressing factor as a power series in $\lambda$. Then we can use the Ward identities to compute the correlation function order by order.
For two-point functions, we thus have the momentum space expansion, 
\begin{align}\label{OpOp}
&    \langle \MOh(p,\bar{p})\MOh(-p,-\bar{p}) \rangle= \sum_{n_1,\bar{n}_1, n_2,{\bar n}_2}  [n_1,\bar{n}_1;n_2,{\bar n}_2] 
   \end{align}
where
\begin{align}\label{term-n}
&  [n_1,\bar{n}_1;n_2,{\bar n}_2] = {(i \frac{\lambda}{\pi})^{n_1+\bar n_1} (-i \frac{\lambda}{\pi})^{n_2+\bar n_2} p^{\bar n_1+\bar n_2} {\bar p}^{ n_1+ n_2} \over {n_1}!{{\bar n}_1}!{n_2}!{{\bar n}_2}!}  \int dx  d\bar{x} \, e^{ipx + i\bar{p}\bar{x}} \, G_{n_1,\bar{n}_1;n_2,{\bar n}_2}(x, 0) 
\nonumber\\
    & G_{n_1,\bar{n}_1;n_2,{\bar n}_2}(x_1,x_2) = \left\langle:{\chi}_{\bar{T}}^{\bar{n}_1}\chi_T^{n_1}\tilde{\MO}_{h,\bar h}(x_1)::\bar{\chi}_{\bar{T}}^{\bar{n}_2}\chi_T^{n_2}\tilde{\MO}_{h,\bar h}(x_2):  \right\rangle, \\
    &  {\bar\chi}_{\bar T}(\bar x)=\int^{\bar{x}} \bar{T}(\bar{x}') \, d\bar{x}',\quad  \chi_T(x)=\int^{{x}} {T}({x}') \, d{x}', \nonumber
\end{align}
and we will use point splitting regularization
\begin{equation}\label{pointsplitting}
    :\chi^n(x)\tilde{\MO}(x):=\oint_{x+\epsilon}\frac{dw}{2\pi i} \frac{1}{w-x}\chi^n(w)\tilde{\MO}(x)-(\log(\epsilon)\text{-divergence})
\end{equation}
In the following, we will first carry out the perturbation calculation up to the second order.  Then we will show that the leading UV contribution can be summed up, and the result agrees perfectly with that of \cite{Cui:2023jrb} obtained from string theory. This result is also consistent with the large momentum limit of the path integral calculation \cite{Aharony:2023dod}.
 
\subsection{Perturbative results}
At the zeroth order,  the two-point function is the same as in the undeformed CFT, which in position space and momentum space are written as 
\begin{align}
G(x,\bar x)_{(0)}=\frac{1}{|x|^{4h}}, \quad  G(p,\bar p)_{(0)}= \frac{\pi\Gamma(1-2h)}{\Gamma(2h)}\z(\frac{|p|}{2}\y)^{4h-2}
\end{align}
where we have chosen $\bar h=h$ for simplicity. 
Using the Ward identity 
\begin{equation}
    \begin{aligned}
        &\langle  T(w_1)\prod_{i=1}^{N}\tilde\MO_i(x_i)\rangle=\z\{\sum_{i=1}^N\z[\frac{h_i}{(w_1-x_i)^2}+{1 \over w_1-x_i }\p_{x_i}\y]\y\}\langle \prod_{i=1}^{N}\tilde\MO_i(x_i)\rangle    \end{aligned}
\end{equation}
we get
\begin{equation}
    \begin{aligned}
        &\langle  \chi_T(w_1)\prod_{i=1}^{N}\tilde\MO_i(x_i)\rangle=\z\{\sum_{i=1}^N\z[-\frac{h_i}{w_1-x_i}+\log(w_1-x_i)\p_{x_i}\y]\y\}\langle \prod_{i=1}^{N}\tilde\MO_i(x_i)\rangle    \end{aligned}
\end{equation}
which will be the building block of the perturbative calculation. 
\subsubsection*{The first order in $\lambda$}
In the linearized two-point function, we have
\begin{equation}
\begin{aligned}
   G_{1,0;0,0}(x_1,x_2)=\z\langle \chi_T(x_1)\tilde{\MO}_{h,\bar{h}}(x_1)\tilde{\MO}_{h,\bar{h}}(x_2)\y\rangle&=\z(-\frac{h}{x_1-x_2}+\log(x_1-x_2)\p_{x_2}\y)\frac{1}{|x_1-x_2|^{4h}}
\end{aligned}
\end{equation}
and similar results for all terms with  $n_1+n_2+\bar n_1+\bar n_2=1$ in \eqref{term-n}. 
 The final result in the position space is given by
 \begin{align}\label{g1x}
G(x,\bar x)_{(1)}  =\frac{2\lambda }{\pi }  F_1(\log |x|^2)\p\bar \p \frac{1}{|x|^{4h}}, \quad  F_1(\theta):=\theta-1
\end{align}
where we have introduced a polynomial of the log function of order $1$ with unit leading coefficient. 
To write the result in the above form, we have moved all the derivatives to the right, and replaced additional powers of $x^1$ by acting $\p$ on the undeformed two-point function.  
From the expression \eqref{g1x}, we can obtain the momentum result by performing the Fourier transformation. 
To express the result in a more suggestive way, we move the derivative to the left, so that each total derivative in  $G(x,\bar x)_{(1)}$ will contribute a factor proportional to $p$ after the Fourier transformation. During the process, $\p$ will hit the polynomial $F_1$ ,which turns $\log x$ into $1/x$. We can again replace $1/x$ by acting $\p$ to the right, which is now a total derivative and can be replaced by $p$.  
 This procedure allows us to organize the momentum space two-point function in the following form, 
\begin{align}\label{g1p}
   G(p,\bar{p})_{(1)}   &=\int dx^2 e^{ipx+i\bar{p}{\bar x}} G(x,\bar x)_{(1)}\nonumber \\
   &=-{2\lambda p\bar p\over \pi} \int dx^2 e^{ipx+i\bar{p}\bar{x}} A_1(\log |x|^2) \frac{1}{|x|^{4h}},\quad  A_1(\theta):=\theta-{(h-1)\over h} 
       \end{align}   
       The polynomial $A_1$ is of the same order as $F_1$, and has the same leading coefficient. The subleading coefficients are different as a result of moving derivatives from the right to the left. 
The advantage of the expression in terms of a polynomial in $\log |x|^2$ is that we can see clearly what the most UV-sensitive term is at each order. 
 The dominant contribution in the UV is the leading term in both the polynomial 
 $F_1$ and $A_1$ at this order. This part is universal and does not depend on the regularization scheme. In fact,  the leading $\log$ in \eqref{g1x} and \eqref{g1p} are compatible with previous results \cite{Cardy:2019qao,Cui:2023jrb,Aharony:2023dod}.

\subsection*{The second order in $\lambda$:}
At the second order, we have $10$ terms, and we need to use the following Ward identities:
\begin{equation}\label{Ward1}
    \begin{aligned}
        &\langle T(w_1) T(w_2) \tilde\MO(x_1){\tilde\MO}(x_2)\rangle\\
        &=\z\{\frac{2}{(w_1-w_2)^2}+\frac{1}{w_1-w_2}\p_{w_2}+\sum_{i=1}^2\z[\frac{h}{(w_1-x_i)^2}+\frac{1}{w_1-x_i} \p_{x_i}\y] \y\} \times \\
        &\quad \z\{ \sum_{j=1}^2\z[\frac{h}{(w_2-x_j)^2}+\frac{1}{w_2-x_j} \p_{x_j}\y] \y\}\langle \tilde\MO(x_1){\tilde\MO}(x_2)\rangle+\frac{c}{2(w_1-w_2)^4}\langle{\tilde\MO}(x_1){\tilde\MO}(x_2)\rangle\\
    \end{aligned}
\end{equation}
and
\begin{equation}\label{Ward2}
    \begin{aligned}
        &\langle T(w_1) \bar{T}(\bar{w}_2) \tilde\MO(x_1){\tilde\MO}(x_2)\rangle\\
        &=\sum_{i=1}^2 \sum_{j=1}^2 \z[\frac{h}{(w_1-x_i)^2}+\frac{1}{w_1-x_i} \p_{x_i}\y]  \times 
          \z[\frac{h}{(\bar{w}_2-\bar{x}_j)^2}+\frac{1}{\bar{w}_2-\bar{x}_j} \p_{\bar{x}_j}\y] \langle \tilde\MO(x_1){\tilde\MO}(x_2)\rangle\\
    \end{aligned}
\end{equation}
The final result in position space is reduced to:
\begin{equation}\label{positionorder2}
\begin{aligned}
G(x, \bar{x})_{(2)} & = \frac{1}{2! } \left(\frac{2 \lambda}{\pi}\right)^{2} F_2(\log |x|^2)\partial_x^2 \partial_{\bar{x}}^2 G(x, \bar{x})_{(0)},\\
F_2(\theta)&:=\theta^2-\frac{\left( 8 h+13\right) }{2(2 h+1)}\theta  + \frac{8 h \left(4 h^2+8 h+5\right)+ (2 h+1) c / 3}{8 h(2 h+1)^2} 
\end{aligned}
\end{equation}
and in the momentum space 
\begin{equation}\label{momentumorder2}
\begin{aligned} G(p, \bar{p})_{(2)} & = {1\over 2!}\int dx^2 e^{ipx+i\bar p\bar x} ({-2\lambda p \bar p  \over \pi})^2 A_2(\log|x|^2)  G(x,\bar{x})_{(0)}\\
A_2(\theta)&=\theta^2 -\frac{8h^2-3h-4}{2h(2h+1)}\theta+ \frac{96h^4-192h^3+2(c-60)h^2+(84+c)h+36}{24h^2(1+2h)^2}
\end{aligned}
\end{equation}
Similar to the linearized case, the polynomials $A_2$ and $F_2$ have the same order and same leading coefficient, which is not affected when we move the derivatives from the right to the left. 
For more details of the calculations, see the appendix \ref{app:2ndOrder}.  

\subsection{A non-perturbative result in the UV}

Similar to the expressions of first order and second order, the $n$-th order correlator can be organized as follows,
\begin{equation}\label{gnx}
G(x, \bar{x})_{(n)}=\frac{(2 \lambda)^n}{\pi^n n!} F_n\left(\theta \right) \partial^n \bar{\partial}^n G(x, \bar{x})_{(0)}, \quad \theta=\log |x|^2
\end{equation}
where $F_n(\theta)$ is a $n$-th order polynomial in $\theta$ with unit leading coefficient:
\begin{equation} \label{FnP}
    F_n(\theta)=\theta^n+c_n^{n-1} \theta^{n-1}+\cdots c_n^0
\end{equation}
At each order, the most dominant contribution at UV in $F(\log|x|^2)$  is given by the leading log term. Summing these up, we can write a formal expression in the position space,
\begin{align}\label{2pfx}
G(x,\bar x)&\sim \int d\theta \delta (\theta -\log|x|^2) e^{ \frac{2\lambda \theta}{\pi}\p_x\p_{\bar{x}}}|x|^{-4h}  
\end{align}
where we have introduced an auxiliary variable $\theta$ to avoid the terms coming from the commutator between $\log|x|^2$ and $\p$. 
The integral over $\theta$ should be taken at the last step of the computation. 

We note the structure of the correlation function \eqref{2pfx} is reminiscent of the star product in non-commutative field theory \cite{Seiberg:1999vs}. In \cite{Shyam:2022iwd}, it was proposed that the $T\bar T$ deformed correlators be obtained by replacing ordinary products with star products, the resulting of which flow equation is similar to that of the undeformed operators \cite{Cardy:2019qao,Kruthoff:2020hsi}. On the other hand, the physical operator we propose in \eqref{OphypQ} leads to the flow equation \eqref{floweq}, which is different from that of the undeformed operators. Nevertheless, it would be interesting to further explore the connection with the star product formulation.

It is interesting to note that the above two-point correlator in position space agrees with the leading-log result of \cite{Hirano:2024eab}, which was obtained by treating the $T\bar{T}$-deformed CFT on the plane as a CFT living on a deformed geometry with nonlocal coordinates.\footnote{See the section 4 of \cite{Hirano:2024eab} for the strategy. The two-point correlation function is given in eq. (5.25), which matches our \eqref{2pfx}. } It is not clear whether the operator discussed in \cite{Hirano:2024eab} is the same as the physical operator introduced in the present paper, and a more careful comparison is needed to fully clarify this point. Nevertheless, we expect the leading-log result \eqref{2pfx} to be universal and thus independent of the precise definition of the operators.

A more compact result can be obtained in the momentum space. At the $n$-th order, we have 
\begin{align}
G(p,\bar{p})_{(n)}&=\int dx^2 e^{ipx+i\bar p\bar x}{(-2\lambda p \bar p )^n \over \pi^nn!} A_n(\log|x|^2)  G(x,\bar{x})_{(0)},\label{Gnp}
\end{align}
where $A_n(\theta)$ is a polynomial of degree $n$ 
\begin{align}
A_n(\theta)&=\theta^n+a_{n,n-1} \theta^{n-1}+\cdots a_{n,0}  
\end{align}
To obtain the above expression from the position space \eqref{gnx}, we need to perform a Fourier transformation and repeat the following procedure:
Move a derivative to the left by using integration by parts, replace total derivatives in the position space by momentum in the Fourier transformation, replace $1/x$ resulting from acting $\p$ on $F_n$ by a derivative to $G_{(0)}$. If the term is already a total derivative, replace the derivative by multiplying by the momentum. If the term is not a total derivative, repeat the procedure. 
The leading log terms in $A_n$ can be summed over, and we get the most UV-sensitive part in the momentum space,
\begin{align}\label{Gn}
\nonumber\langle \MO_{h,\bar{h}}(p,\bar{p})\MO_{h,\bar{h}}(-p,-\bar{p}) \rangle  &\sim  \int d x^2 e^{i p x+i \bar{p} \bar{x}} \, |x|^{-\frac{4\lambda}{\pi}  p\bar {p}} \frac{1}{|x|^{4 h}} \\ &= \frac{\pi \Gamma\left(1-2 h_\lambda\right)}{\Gamma\left(2 h_\lambda\right)}\left(\frac{|p|}{2}\right)^{4 h_\lambda-2},\quad 
h_{\lambda} = h+ \frac{\lambda}{\pi} p\bar{p}
\end{align}
which looks like a CFT two-point function with shifted conformal weight $h_{\lambda}$. A similar procedure can be performed for the $N$-point correlation functions, and we can show
\begin{equation}
    \langle\prod^{N}_{i=1}\MO_{h_i,\bar{h}_i}(p_i,\bar{p}_i)\rangle\sim\mathcal{F}_{p_i,\bar{p}_i}\z[\prod_{j=1}^N\prod_{k\neq j,k=1}^N|(x_j-x_k)^{\frac{\lambda\bar{p}_jp_k}{\pi}}|^2\langle\prod^N_{i=1}\tilde{\MO}_{h_i,\bar{h}_i}(x_i,\bar{x}_i)\rangle\y]
\end{equation}
where again we have only kept the leading-log terms. The leading UV result \eqref{Gn}  agrees perfectly with the result obtained from holographic calculation in the TsT/$T\bar T$ correspondence \cite{Cui:2023jrb}, and the large momentum result of $T\bar T$ calculation in the path integral formalism \cite{Aharony:2023dod}. As discussed in \cite{Cui:2023jrb}, this result also satisfies the Callan-Symanzik equation derived in \cite{Cardy:2019qao}.

It is interesting to note that in the large momentum limit, the two point function goes like
\begin{equation}
    \langle \MOh(p,\bar{p})\MOh(-p,-\bar{p})\rangle\sim |p|^{-{4\over\pi}\lambda |p|^2}
\end{equation}
which is exponentially suppressed for $\lambda>0$, and exponentially enhanced for $\lambda<0$. In both cases, the two point function deviates from the typically polynomial suppression in standard local quantum field theory, and hence  indicates that the $T\bar T$ deformed CFT is non-local. For $\lambda>0$, the exponentially suppressed behavior suggests that the theory is UV-soft but nevertheless non-local smearing, and the combination $\lambda |p|^2$ suggests a fundamental scale  $\sqrt{\lambda}$. These properties are similar to string theory, compatible with \cite{Dubovsky:2012wk}.  For $\lambda<0$,
the exponential growth indicates UV catastrophe, and is compatible with the fact $T\bar T$ deformed CFT has complex spectrum at high energy on the cylinder.

\section{Summary and Discussion}

In this paper, we introduce a novel class of  operators in $T\bar{T}$-deformed CFTs, which we refer to as physical operators. These operators bridge two previously studied bases: the undeformed operators, which are defined via the standard time evolution of seed CFT operators \eqref{undeformedO}, and the dressed operators, which rigidly satisfy the covariant flow equation \eqref{fO}. At the classical level, the physical operators can be expressed in two equivalent forms. First, they admit a local algebraic expression in terms of the undeformed operator \eqref{physicalOp}. Second, they can be rewritten as an integral expression over nonlocal coordinates mapping to the dressed operator, given by \eqref{Ophys2} in position space and \eqref{TTbardress} in momentum space.

By framing the $T\bar{T}$ deformation as a canonical transformation following \cite{Kruthoff:2020hsi} and solving the flow equation \eqref{flowedeq}, we obtain explicit solution of dressed operators on the plane \eqref{Flowedoperator}. 
Solving the flow equations for the stress tensor yields the explicit dressed components \eqref{flowed HLR}, generating two commuting Virasoro algebras \eqref{Witt} that are perfectly compatible with the large-radius limit of the cylinder results \cite{Guica:2022gts}.
The ward identities between dressed  operators and dressed stress tensor do not depend on the deformation parameter, which means that primary operators in the seed CFT remain primary under the dressed conformal symmetry.

At the quantum level, we assume that these dressed operators still exist and inherit the Ward identities of the dressed stress tensor and preserve the functional form of the correlation functions found in the undeformed CFT$_2$. Taking the dressed operator as the building block, we define the physical operator at the quantum level by promoting its classical integral relation to the dressed operators \eqref{OphysQ} and \eqref{OphypQ}. To rigorously handle the inherent short-distance singularities, we employ a point-splitting normal-ordering prescription in Euclidean space \eqref{pointsplitting}. It is interesting to understand how to fully promote the classical relationship between the undeformed operator and the dressed operator \eqref{Flowedoperator}, or the physical operator \eqref{physicalOp}, to a complete quantum relation.  Nevertheless the computation of correlation function does not depend on these relations.

The structural relation between the physical operator and dressed operator allows us to utilize the nonlocal conformal symmetry to compute correlation functions of physical operators. By expanding the Wilson-line-like dressing factor of the momentum space operator \eqref{OE} in powers of $\lambda$, we carried out a systematic perturbative calculation of the two-point function. By extracting and summing the most UV-sensitive, leading-logarithmic terms across all orders, we obtained a compact, non-perturbative expression valid in the large momentum limit \eqref{Gn}. This formulation exhibits an effective, momentum-dependent shifted conformal weight $h_\lambda = h + \frac{\lambda}{\pi}p\bar{p}$. This specific large-momentum behavior exactly matches both the field-theoretic path-integral limits derived in \cite{Aharony:2023dod} and the string theory predictions via the TsT/$T\bar{T}$ correspondence \cite{Cui:2023jrb}.

 We conclude by pointing out that the physical operator we introduce has the following features:
\begin{itemize}
    \item It possesses a purely local classical definition \eqref{physicalOp} constructed strictly from undeformed fields and the physical stress tensor.
    \item It perfectly agrees, at the classical level, with the alternative target-space operators defined by coupling the original CFT to topological gravity as introduced in \cite{Aharony:2023dod}.
    \item It manifests naturally in the nonlocal conformal symmetry framework via integral relations to dressed operators in position \eqref{Ophys2} and momentum \eqref{TTbardress} spaces, facilitating systematic correlation function computations.
 \item It has a fixed spin, as can be seen from the Ward identity with the boost generator \eqref{boostcharge1} and \eqref{boostchargeM}. 
 \item As commented below \eqref{boostcharge1}, the physical operator is the unique operator within the ansatz \eqref{ansatz} which satisfy two conditions: the Ward identity under boost transformation \eqref{boostcharge1} and the flow is a total derivative. 
\item The large momentum limit of the two-point function indicates that the $T\bar T$ deformed CFT is nonlocal.
\end{itemize}

So far, we do not have a first-principle derivation of the physical operator we introduced in this paper. It is possible that there exists other interesting operator basis. 
 Several interesting directions are left for future investigation: to study the operator on the cylinder, to explore possible alternative operator choices, to establish the quantum relation between the undeformed operator and the physical operator, to obtain a compact result of correlation functions without taking the large momentum limit, and to systematically explore the structure of causality and locality.  


\section*{Acknowledgements}
We thank Luis Apolo, Monica Guica, Yunfeng Jiang, Wen-Xin Lai, Huajia Wang, Juntao Wang, Xianjin Xie, and Changhe Yang for useful discussions. The work is supported by the NSFC special fund for theoretical physics No. 12447108 and the national key research and development program of China No. 2020YFA0713000. W.S. would like to thank the Annual Conference of the Simons Foundation Collaboration on Celestial Holography, where partial results of this work were presented and valuable feedback was received. Z.D. would like to thank the support of the PhD student fund for short-term overseas visits and the Simons Foundation Collaboration on Celestial Holography. L.C. would also like to thank the support of the Shuimu Tsinghua Scholar Program of Tsinghua University.

 \appendix

 \section{The Poisson brackets}\label{appdxA}
  In all our calculations, we have replaced the quantum Dirac bracket with the classical Poisson bracket  by the standard rule:
  \begin{equation}
      \big[F, \,G \big] \rightarrow i \big\{F, \, G\big\}
  \end{equation}
  where $F, G$ are operators at the equal Lorentz time slice. In two-dimensional classical field theory, $F, G$  should be understood as functionals of canonical variables $\phi(t,y), \, \pi(t, y)$.  Poisson brackets are given by:
\begin{equation}
\begin{aligned}\label{Poisson}
    &\{F(y),\, G(y')\}=\z(\frac{\p F(y)}{\p \phi(y)}\frac{\p G(y')}{\p \pi(y')}-\frac{\p F(y)}{\p \pi(y)}\frac{\p G(y')}{\p \phi(y')}\y)\delta(y-y')\\
    &+\z(\frac{\p F(y)}{\p \p_y\phi(y)}\frac{\p G(y')}{\p \pi(y')}+\frac{\p F(y)}{\p \pi(y)}\frac{\p G(y')}{\p \p_{y'}\phi(y')}\y)\p_y\delta(y-y')
\end{aligned}
\end{equation}
In this appendix, we will list some useful Poisson brackets that are helpful in deriving the results in the main text. 

\subsection{Poisson brackets for $T\bar{T}$ deformation: }
Let us first provide some details of Poisson brackets between different components of the stress tensor \eqref{HP}. 
 From the definition \eqref{Poisson}, we have:
\begin{equation}
    \begin{aligned}
        \left\{\mathcal{H}(y), \mathcal{H}\left(y^{\prime}\right)\right\} &= \delta(y - y') \left(\frac{\p \mathcal{H}(y)}{\p\phi(y)} \frac{\p\mathcal{H}(y')}{\p\pi(y')}-\frac{\p \mathcal{H}(y)}{\p \pi(y)} \frac{\p \mathcal{H}(y')}{\p \phi(y')}\right) \\
        &+\frac{\p \MH(y)}{\p \p_y \phi(y)}\frac{\p \MH(y')}{\p \pi (y')}\p_y\delta(y-y')-\frac{\p \MH(y)}{\p \pi (y)}\frac{\p \MH(y')}{\p \p_{y'} \phi(y')}\p_{y'}\delta(y-y')\\
        &=\left(\mathcal{P}(y)+\mathcal{P}\left(y^{\prime}\right)\right) \partial_y \delta\left(y-y^{\prime}\right)\\
        \{\MH(y),\MP(y')\} &=\frac{\p \MH(y)}{\p \phi(y)}\p_{y'}\phi(y')\delta(y-y')- \bigg[\frac{\p\MH(y)}{\p\p_y\phi(y)}\p_{y'}\phi(y') + \frac{\p\MH(y)}{\p\pi(y)}\pi(y') \bigg] \p_{y'}\delta(y-y')\\
        &=\p_y \MH(y)\delta(y-y')-(T_{yy}(y)+\MH(y))\p_{y'}\delta(y-y')\\
        &=(T_{yy}(y)+\MH(y'))\p_y\delta(y-y')
    \end{aligned}
\end{equation}
A similar calculation yields the result of $\left\{\mathcal{P}(y), \mathcal{P}\left(y^{\prime}\right)\right\}$ in \eqref{HP}.

In order to solve the flow equation, we need to consider the Poisson bracket for the evolution operator of $T\bar T$ deformation
\begin{equation}
    \mathcal{X}_{T\bar{T}}=\int dy dy'G(y-y')\MH(y)\MP(y')
    \end{equation}
    which appears in the covariant derivative \eqref{feqT}.
Two basic Poisson brackets are:
\begin{equation}
\begin{aligned}
& \left\{\mathcal{X}_{T \bar{T}}, \, \mathcal{H}(y)\right\}=\mathcal{H}\left(T_{y y}+\mathcal{H}\right)-2 \mathcal{P}^2+\int d y_1 G\left(y_1-y\right)\bigg(\mathcal{P}\left(y_1\right) \partial_y \mathcal{P}(y)-\mathcal{H}\left(y_1\right) \partial_y \mathcal{H}(y)\bigg) \\
& \left\{\mathcal{X}_{T \bar{T}}, \, \mathcal{P}(y)\right\}=\mathcal{P}(y)\left(\mathcal{H}-T_{y y}\right)+\int d y_1 G\left(y_1-y\right)\bigg(\mathcal{P}\left(y_1\right) \partial_y T_{y y}(y)-H\left(y_1\right) \partial_y \mathcal{P}(y)\bigg)
\end{aligned}
\end{equation}
Here we have used the integration by parts and throw away the term of the form
\begin{equation}
    G(y-y')\mathcal{H}_L(y)\mathcal{H}_R(y)|_{y=\pm \infty}
\end{equation} which is allowed as long as 
 the stress tensor $\mathcal{H}_{L,R}$ vanish at infinity as required in \eqref{vanish}.
Furthermore, we can get
\begin{equation}
    \mathcal{D}_\lambda \eta_0 = 2 \, \bar\eta_0 \, \mathcal{H}_{L} + \lambda  \z(\eta_0-{\bar\eta}_0\y) \,\mathcal{O}_{T \bar{T}},\quad \mathcal{D}_\lambda {\bar\eta}_0 = 2 \, \eta_0 \, \mathcal{H}_{R} + \lambda \z({\bar \eta}_0-{\eta}_0\y) \,\mathcal{O}_{T \bar{T}}
\end{equation}
Here we also use integration by parts and assume that $\eta_0$ and $\bar{\eta}_0$ remain finite at infinity.
\subsection{Poisson brackets for operators}\label{A2}
We can also have the Poisson brackets between the stress tensor and the undeformed operator $\MOh^{(0)}$
\begin{equation}\label{HO}
\begin{aligned}
\left\{\mathcal{H}_L\left(y_1\right), \mathcal{O}_{h, \bar{h}}^{(0)}\left(y_2\right)\right\}&=\left\{H_L, \mathcal{O}_{h, \bar{h}}^{(0)}\left(y_2\right)\right\} \delta\left(y_1-y_2\right) \\
& \quad + \left(h \frac{\partial \mathcal{H}_L\left(y_2\right)}{\partial \mathcal{H}_L^{(0)}}-\bar{h} \frac{\partial \mathcal{H}_L\left(y_2\right)}{\partial \mathcal{H}_R^{(0)}}\right) \mathcal{O}_{h, \bar{h}}^{(0)}\left(y_2\right) \partial_{y_1} \delta\left(y_1-y_2\right) \\
\left\{\mathcal{H}_R\left(y_1\right), \mathcal{O}_{h, \bar{h}}^{(0)}\left(y_2\right)\right\} &=\left\{H_R, \mathcal{O}_{h, \bar{h}}^{(0)}\left(y_2\right)\right\} \delta\left(y_1-y_2\right) \\
&\quad +\left(h \frac{\partial \mathcal{H}_R\left(y_2\right)}{\partial \mathcal{H}_L^{(0)}}-\bar{h} \frac{\partial \mathcal{H}_R\left(y_2\right)}{\partial \mathcal{H}_R^{(0)}}\right) \mathcal{O}_{h, \bar{h}}^{(0)}\left(y_2\right) \partial_{y_1} \delta\left(y_1-y_2\right)
\end{aligned}
\end{equation}
Now we consider the factor of the physical operator 
\begin{equation}
	F=(1+\lambda^2\OT),\quad W_{L}=(1+2\lambda\HR) \quad ,W_{R}=(1+2\lambda\HL)
\end{equation}
Their Poisson brackets with the stress tensor are:
\begin{equation}
	\begin{aligned}
		&\{\HL(y_1),F^a(y_2)\}=\{H_L,F^a(y_2)\}\delta(y_1-y_2)+\frac{8a\lambda^2\HL\HR F^a(y_2)}{(1+2\lambda \MH)^2}\p_{y_1}\delta(y_1-y_2)\\
		&\{\HR(y_1),F^a(y_2)\}=\{H_R,F^a(y_2)\}\delta(y_1-y_2)-\frac{8a\lambda^2\HL\HR F^a(y_2)}{(1+2\lambda \MH)^2}\p_{y_1}\delta(y_1-y_2)\\
		&\{\HL(y_1),W_{L}^b(y_2)\}=\{H_L,W_{L}^b(y_2)\}\delta(y_1-y_2)\\
		&\{\HR(y_1),W_{L}^b(y_2)\}=\{H_R,W_{L}^b(y_2)\}\delta(y_1-y_2)-\frac{4b\lambda\HR W_{L}^b(y_2)}{1+2\lambda\MH}\p_{y_1}\delta(y_1-y_2)\\
		&\{\HL(y_1),W_{R}^c(y_2)\}=\{H_L,W_{R}^c(y_2)\}\delta(y_1-y_2)+\frac{4c\lambda\HL W_{R}^c(y_2)}{1+2\lambda\MH}\p_{y_1}\delta(y_1-y_2)\\
		&\{\HR(y_1),W_{R}^c(y_2)\}=\{H_R,W_{R}^c(y_2)\}\delta(y_1-y_2)
	\end{aligned}
\end{equation}
Then, we can get the covariant derivative of  \eqref{Cphyflow} in terms of these ingredients.
\subsection{Calculation of $\MD_\lambda \tilde{\MO}_{h,\bar{h}}^{(n)}$}
Here, we present some key steps for the proof of $\MD_\lambda \tilde{\MO}_{h,\bar{h}} = 0 $.  To this end,  we compute the derivative of $ \tilde{\MO}_{h,\bar{h}}^{(n)}$, which is:
\begin{equation}
	\begin{aligned}
		\MD_\lambda\tMOh^{(n)}=(-1)^n \sum_{m=0}^n \frac{1}{(n-m)!m!}\p_u^m\p_v^{n-m}Q^{(n)}_{m}
	\end{aligned}
\end{equation}
where,
\begin{equation}
	\begin{aligned}
		Q^{(n)}_m&=(\delta\hat{u})^{m}(\delta\hat{v})^{n-m}\MOh\z(\frac{m\MD_\lambda\delta\hat{u}}{\delta\hat{u}}+\frac{(n-m)\MD_\lambda\delta\hat{v}}{\delta\hat{v}}\y)\\
        &+\frac{1}{\lambda}(\delta\hat{u})^{m}(\delta\hat{v})^{n-m}(\p_u(\delta\hat{u}\MOh)+\p_v(\delta\hat{v}\MOh))\\
		&=\frac{1}{\lambda}\p_u\z((\delta\hat{u})^{m+1}(\delta\hat{v})^{n-m}\MOh\y)+\frac{1}{\lambda}\p_v\z((\delta\hat{u})^{m}(\delta\hat{v})^{n-m+1}\MOh\y)+\frac{n}{\lambda}(\delta\hat{u})^{m}(\delta\hat{v})^{n-m}\MOh
	\end{aligned}
\end{equation}
Then, notice the following identity:
\begin{equation}
	\begin{aligned}
		&\frac{(-1)^n}{\lambda} \sum_{m=0}^n \frac{1}{(n-m)!m!}\p_u^{m+1}\p_v^{n-m}((\delta\hat{u})^{m+1}(\delta\hat{v})^{n-m}\MOh)+ \\
        &\qquad \frac{(-1)^n}{\lambda} \sum_{m=0}^n \frac{1}{(n-m)!m!}\p_u^m\p_v^{n-m+1}((\delta\hat{u})^{m}(\delta\hat{v})^{n-m+1}\MOh)\\
		&=\frac{(-1)^n}{\lambda} \sum_{m=0}^{n+1} \frac{m}{(n-m+1)!m!}\p_u^{m}\p_v^{n-m+1}((\delta\hat{u})^{m}(\delta\hat{v})^{n-m+1}\MOh)+ \\
        & \qquad \frac{(-1)^n}{\lambda} \sum_{m=0}^{n+1} \frac{n+1-m}{(n-m+1)!m!}\p_u^m\p_v^{n-m+1}((\delta\hat{u})^{m}(\delta\hat{v})^{n-m+1}\MOh)\\
		&=-\frac{n+1}{\lambda}\tMOh^{(n+1)},
	\end{aligned}
\end{equation}
we thus have:
\begin{equation}
	\MD_\lambda\tMOh^{(n)}=\frac{n}{\lambda}\tMOh^{(n)}-\frac{n+1}{\lambda}\tMOh^{(n+1)}
\end{equation}
By using this recursive relation,  we can get the solution of the dressed operator \eqref{Flowedoperator}.

\section{The second order perturbative calculations of correlation functions}\label{app:2ndOrder}

In this appendix, we will spell out the details of the second-order perturbative calculations of the two-point functions. By using the Ward identities \eqref{Ward1} and \eqref{Ward2} and  considering the normal ordering, we have:
\begin{equation}
    \begin{aligned}
                &\langle \chi_T(x_1) \chi_T(x_2)\tilde{\MO}(x_1)\tilde{\MO}(x_2)\rangle=\Big[-\frac{h^2}{(x_1-x_2)^2}+\frac{4h\log(x_1-x_2)}{(x_1-x_2)^2}-\frac{c}{12(w_1-w_2)^2}\\
        &\qquad -\frac{(1+h)\log(x_1-x_2)}{x_1-x_2}\left(\p_{x_1}-\p_{x_2}\right) +\log(x_2-x_1)^2\p_{x_1}\p_{x_2}\Big]\langle \tilde{\MO}(x_1)\tilde{\MO}(x_2)\rangle, \\
                &\langle \chi_T(x_1) \chi_T(x_1)\tilde{\MO}(x_1)\tilde{\MO}(x_2)\rangle=\Big[\frac{h(1+h)}{(x_1-x_2)^2}-\frac{2h\log(x_1-x_2)}{(x_1-x_2)^2}  -\frac{2(1+h)\log(x_1-x_2)}{x_1-x_2}\p_{x_2}\\
        &\qquad +\log(x_1-x_2)^2\p_{x_2}^2\Big]\langle \tilde{\MO}(x_1)\tilde{\MO}(x_2)\rangle, \\
          &\langle \chi_T(x_1) \bar{\chi}_T(\bar{x}_1)\tilde{\MO}(x_1)\tilde{\MO}(x_2)\rangle=\Big[\frac{h^2}{|x_1-x_2|^2}  -\frac{h\log(x_1-x_2)}{\bar{x}_1-\bar{x}_2}\p_{x_2}-\frac{h\log(\bar{x}_1-\bar{x}_2)}{x_1-x_2}\p_{\bar{x}_2} \\
          & \qquad +\log(x_1-x_2)\log(\bar{x}_1-\bar{x}_2)\p_{x_2}\p_{\bar{x}_2}\Big]\langle \tilde{\MO}(x_1)\tilde{\MO}(x_2)\rangle , \\
      &\langle \chi_T(x_1) \bar{\chi}_T(\bar{x}_2)\tilde{\MO}(x_1)\tilde{\MO}(x_2)\rangle=\Big[-\frac{h^2}{|x_1-x_2|^2}  -\frac{h\log(\bar{x}_2-\bar{x}_1)}{x_1-x_2}\p_{\bar{x}_1}-\frac{h\log(x_1-x_2)}{\bar{x}_2-\bar{x}_1}\p_{x_2} \\
      &\qquad +\log(x_1-x_2)\log(\bar{x}_2-\bar{x}_1)\p_{x_2}\p_{\bar{x}_1}\Big]\langle \tilde{\MO}(x_1)\tilde{\MO}(x_2)\rangle
    \end{aligned}
\end{equation}
By summing them up, we can obtain the second-order contribution:
\begin{equation}
    \begin{aligned}
        & G(p,\bar{p})_{(2)}  =\frac{\lambda^{2}}{\pi^{2}} \int dx^2 \, e^{ipx +i\bar{p}\bar{x}} \Bigg\{ {4 p\bar{p}}\z(-\frac{h^2}{|x|^2}-\frac{h\log x}{\bar{x}}\p_x-\frac{h\log \bar{x}}{x}\p_{\bar{x}}-\log x\log\bar{x}\p_x\p_{\bar{x}}\y) \\
        & \quad +{p^2} \z(-\frac{c}{12\bar{x}^2}-\frac{h(1+2h)}{\bar{x}^2}+\frac{6h\log \bar{x}}{\bar{x}^2}-\frac{4(1+h)\log \bar{x}}{\bar{x}}\p_{\bar{x}}-2(\log \bar{x})^2\p_{\bar{x}}^2\y)\\
        &\quad + {\bar{p}^2} \Big(-\frac{c}{12x^2}-\frac{h(1+2h)}{x^2}+\frac{6h\log x}{x^2}-\frac{4(1+h)\log x}{x}\p_x 
        -2(\log x)^2\p_x^2 \Big)  \Bigg\} \, \left\langle \tilde{\MO}(x)\tilde{\MO}(0) \right\rangle\\
    \end{aligned}
\end{equation}
Then, we perform the integration by parts to move all the derivatives to the right, and the result can be reduced further as:
\begin{equation}
\begin{aligned}
     G(p,\bar{p})_{(2)}  &= \int dx^2 e^{ipx +i\bar{p}\bar{x}} \, G(x, \bar{x})_{(2)} \\
G(x, \bar{x})_{(2)} & = \frac{1}{2! } \left(\frac{2 \lambda}{\pi}\right)^{2} F_2(\log |x|^2) \, \partial_x^2 \partial_{\bar{x}}^2 \, G(x, \bar{x})_{(0)},\\
F_2(\theta)&=\theta^2-\frac{\left( 8 h+13\right) }{2(2 h+1)}\theta  + \frac{8 h \left(4 h^2+8 h+5\right)+ (2 h+1) c / 3}{8 h(2 h+1)^2} 
\end{aligned}
\end{equation}
where $G(x,\bar{x})_{(0)} = {|x|^{-4h}}$ is the leading order two-point function. Equivalently, we can move all the derivatives to the left, so that the differential operator becomes the powers of momentum:
\begin{equation}
\begin{aligned} G(p, \bar{p})_{(2)} & = {1\over 2!}\int dx^2 e^{ipx+i\bar p\bar x} \, \left({-2\lambda p \bar p  \over \pi}\right)^2 A_2(\log|x|^2)  \, G(x,\bar{x})_{(0)}\\
A_2(\theta)&=\theta^2 -\frac{8h^2-3h-4}{2h(2h+1)}\theta+ \frac{96h^4-192h^3+2(c-60)h^2+(84+c)h+36}{24h^2(1+2h)^2}
\end{aligned}
\end{equation}

\bibliographystyle{JHEP}
\bibliography{ref.bib}

\end{document}

%% file: preamble.tex
\usepackage{jheppub}
\usepackage{mathtools}
\usepackage{float}
\usepackage{tikz-cd}

\date{}

\newcommand{\OT}{\mathcal{O}_{T\bar{T}}}

\newcommand{\HL}{\mathcal{H}_L}
\newcommand{\HR}{\mathcal{H}_R}
\newcommand{\MH}{\mathcal{H}}
\newcommand{\MP}{\mathcal{P}}
\newcommand{\MD}{\mathcal{D}}
\newcommand{\MO}{\mathcal{O}}

\newcommand{\MOh}{\mathcal{O}_{h,\bar{h}}}

 \usepackage[normalem]{ulem}
\usepackage[textsize=scriptsize,textwidth=2.5cm]{todonotes}

\let\exporig\exp
\usepackage{physics}
\usepackage{enumitem}
\let\exp\exporig

\affiliation[\ensuremath{\gamma}]{Yau Mathematical Sciences Center, Tsinghua University, Beijing 100084, China}
\affiliation[\ensuremath{\tau}]{Department of Mathematical Sciences, Tsinghua University, Beijing 100084, China}
\affiliation[\ensuremath{\sigma}]{Center for the Fundamental Laws of Nature, Harvard University, Cambridge, MA, USA}

\affiliation[\ensuremath{\eta}]{Peng Huanwu Center for Fundamental Theory, Hefei, Anhui 230026, China}

\usepackage{xspace}